# Opt: A Domain Specific Language for Non-linear Least Squares Optimization in Graphics and Imaging

ZACHARY DEVITO, Facebook Research
MICHAEL MARA, Stanford University
MICHAEL ZOLLHÖFER, Max-Planck-Institute for Informatics
GILBERT BERNSTEIN, Stanford University
JONATHAN RAGAN-KELLEY, UC Berkeley
CHRISTIAN THEOBALT, Max-Planck-Institute for Informatics
PAT HANRAHAN, Stanford University
MATTHEW FISHER, Adobe Research
MATTHIAS NIESSNER, Technical University of Munich

Many graphics and vision problems can be expressed as non-linear least squares optimizations of objective functions over visual data, such as images and meshes. The mathematical descriptions of these functions are extremely concise, but their implementation in real code is tedious, especially when optimized for real-time performance on modern GPUs in interactive applications. In this work, we propose a new language, Opt[1], for writing these objective functions over image- or graph-structured unknowns concisely and at a high level. Our compiler automatically transforms these specifications into state-of-the-art GPU solvers based on Gauss-Newton or Levenberg-Marquardt methods. Opt can generate different variations of the solver, so users can easily explore tradeoffs in numerical precision, matrix-free methods, and solver approaches.

In our results, we implement a variety of real-world graphics and vision applications. Their energy functions are expressible in tens of lines of code, and produce highly-optimized GPU solver implementations. These solvers are competitive in performance with the best published hand-tuned, application-specific GPU solvers, and orders of magnitude beyond a general-purpose auto-generated solver.

## 1 INTRODUCTION

Many problems in graphics and vision can be concisely formulated as least squares optimizations on images, meshes, or graphs. For example, Poisson image editing, shape-from-shading, and as-rigid-as-possible warping have all been formulated as non-linear least squares optimizations, allowing them to be described tersely as energy functions over pixels or vertices [47, 50, 68].

In many of these applications, high performance is critical for interactive feedback, requiring efficient parallel or GPU-based solvers [11, 30, 54, 68, 73]. These solvers require optimizations that are not expressible in generic sparse linear algebra software. For example, these solvers are *matrix-free*; that is, they compute matrix values on-the-fly rather than loading data from matrices *materialized* (i.e., stored) in memory. They also implicitly represent the sparse connectivity of the matrices based on the structure of images or graphs, rather than store it explicitly in memory. This approach can also be applied to solvers that use the Levenberg-Marquardt algorithm [36, 39].

However, the efficiency of these solvers comes at enormous implementation cost: the simple energy function must be manually transformed into a complex product of partial derivative matrices (e.g., $J^T F$ and $J^T Jp$). Furthermore, the code tightly intertwines the calculation of partial derivatives with operations performed by the solver. Finally, this code has to be written by hand in GPU kernels; the result is hundreds of lines of highly-tuned CUDA code which is hard to maintain and modify.

This paper presents a new language, Opt, which makes this type of high performance-optimization accessible to a wider community of graphics and vision practitioners. Programmers write high-level sum-of-squares energy functions over pixels or graphs, such as the example shown in Fig. 1 (left) for as-rigid-as-possible image warping. Our compiler can transform these energies into efficient GPU routines which compute products of the derivatives (e.g., $J^T J$ or $J^T Jp$). We provide a suite of solvers that use these routines to apply either Gauss-Newton (**GN**) or Levenberg-Marquardt (**LM**) methods. The solvers use a parallel preconditioned conjugate gradient inner loop based on either matrix-free or materialized approaches. The resulting code can be used within both matrix-free and materialized solvers.

Our system is able to achieve this due to four key ideas. First, we provide an optimization framework that separates the details of a particular energy from the details of the GN or LM solver approach. The framework is general enough to allow both matrix-free and materialized implementations. Second, our language provides key abstractions for representing energies at a high-level. Unknowns and other data are arranged on 2D or 3D grids, meshes, or general graphs. Energies are defined over these domains and access data through *stencil* patterns (fixed-size and shift-invariant local windows). Third, our compiler exploits the regularity of stencils and graphs to automatically generate efficient GPU routines that can compute products of derivatives such as $\mathbf{J}^T \mathbf{J}$. Derivative terms required by these routines are created using hybrid symbolic-automatic differentiation based on a simplified version of the D⋆ algorithm [26]. Finally, we use a specialized code generator to emit efficient GPU code for the derivative terms and use metaprogramming to weave the solver code with the generated routines to avoid runtime overhead.

[1]Opt is open sourced and publicly available under http://optlang.org.

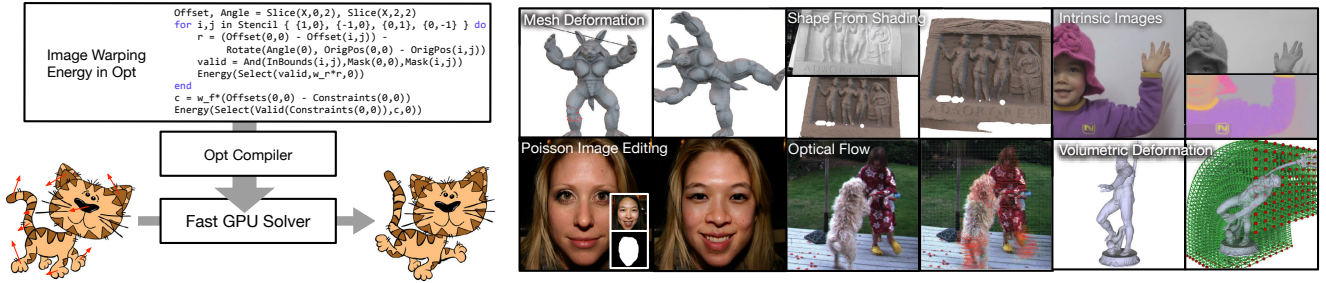

Fig. 1. From a high-level description of an energy, Opt produces high-performance GPU-based optimizers for many graphics problems.

Our method provides significantly better performance than traditional general-purpose solver libraries and matches state-of-the-art custom applications. It is easy to change details in the solver (GN vs. LM), the numeric precision (float vs. double), and matrix storage (matrix-free vs. materialized) without rewriting the energies or solvers. Programmers can quickly figure out the best settings for a particular problem depending on the need for numerical stability and available computational resources.

In particular, we present the following contributions:

- We propose a high-level programming model for defining energies over image and graph domains.
- We introduce a generic framework for Gauss-Newton and Levenberg-Marquardt optimization on GPUs that is capable of abstracting the efficient matrix-free methods used in state-of-the-art application-specific solvers.
- We provide algorithms based on symbolic differentiation that exploit the regularity of energies defined on images and graphs to produce efficient GPU routines that plug into our optimization framework. Our optimizations produce code competitive with hand-written routines.
- We implement a variety of state-of-the-art graphics problems, including mesh/image deformations, smoothing, and shape-from-shading refinement using Opt. We provide an evaluation that shows that our implementations outperform state-of-the-art application-specific solvers and are up to two orders of magnitude faster than the CPU-based Ceres solver [1].
- We show how Opt's abstraction allows the flexible generation of many solver variants for these applications that

explore tradeoffs in GN vs. LM, single vs. double precision, matrix-free vs. materialized, and even hybrid solvers.

## 2 BACKGROUND

*Non-linear Least Squares Optimization.* Optimization methods are used in the graphics and vision community to solve a wide range of problems. We specifically focus on *unconstrained non-linear least squares* optimizations [6], where a solver minimizes an energy function that is expressed as a sum of squared *residual* terms: $E(\mathbf{x}) = \sum_{r=1}^{R} \left[ f_r(\mathbf{x}) \right]^2$. The residuals $f_r(\mathbf{x})$ are generic functions, making the problems potentially non-linear and non-convex [9]. There has been an extensive effort in the literature to solve these problems with a large variety of numerical optimization approaches [13, 14, 21, 33, 43, 44, 66]. Gauss-Newton and Levenberg-Marquardt [36, 39] are two common methods for solving problems in computer graphics and vision.

GN and LM are specifically tailored towards these kind of problems. Their second-order optimization approach has been shown to be well-suited for the solution of a large variety of problems [38, 67], and has also been successfully applied in the context of real-time optimization [68, 73]. If the non-linear energy is convex, then GN and LM will converge to the global minimum; otherwise they will converge to some local minimum. Furthermore, GN and LM internally solve a linear system. While these systems can generally be solved with direct methods, our solvers need to scale to large problem sizes and run on massively parallel GPUs; hence, we implement GN/LM with a preconditioned conjugate gradient (PCG) [45] in the inner loop.

In our current implementation, we focus on GN and LM rather than other variants such as L-BFGS [45], since they reflect the approaches used in state-of-the-art hand-written GPU implementations, allowing us to compare our performance to existing solvers directly. However, we believe our approach can be generalized in future work to support such backends.

*Application-specific GPU Solvers.* Application-specific Gauss-Newton solvers written for GPUs have been frequently used in the last two years. Wu et al. [68] use a blocked version of GN to refine depth from RGB-D data using shape-from-shading. Zollhöfer et al. [73] minimize an as-rigid-as-possible energy [50] on a mesh as part of a framework for real-time non-rigid reconstruction. Zollhöfer et al. [72] use a similar solver to enforce shading constraints on a volumetric signed-distance field in order to refine over-smoothed geometry with RGB data. Thies et al. [54–56] transfer local facial expressions between people in a video by optimizing photo-consistency between the video and synthesized output. Dai et al. [11] solve a global bundle adjustment problem to achieve real-time rates for globally-consistent 3D reconstruction, and Innmann et al. [30] optimize the a warp field for of non-rigid surface reconstruction.

These solvers achieve high-performance by working *matrix-free* on the problem domain. That is, during the PCG step, they never form, or *materialize* the entire Jacobian **J** of the energy. Instead, they compute it on demand, for instance by reading neighboring pixels to compute the derivative of a regularization energy. Performance improves in two ways: first, they do not explicitly store and load sparse matrix connectivity; rather, this is implied by pixel relationships or meshes. Second, reconstructing terms is often faster than storing them, since the size of the problem data is *smaller* than the full matrix implied by the energy. Unfortunately, these application-specific solvers are tedious to write because they mix code that calculates complicated matrix products with partial derivatives based on the energy.

*High-level Solvers.* Higher-level solvers such as CVX [19, 20], OpenOF [64], or ProxImaL [28] work directly from an energy specified in a domain-specific language. CVX uses disciplined programming to ensure that modeled energy functions are convex, then constructs a specialized solver for the given type of convex problem. Ceres [1] uses template meta-programming and operator overloading to solve non-linear least squares problems on the CPU using backwards auto-differentiation. Unlike Opt, these solvers do not generate efficient GPU implementations and only work with materialized matrices. OpenOF does run on GPUs, but uses materialized sparse matrices [64]. In contrast, Opt's abstraction allows solvers to use either matrix-free or materialized approaches; we can even provide hybrids where only part of the energy is materialized. Matrix-free approaches can be significantly faster than explicit matrices due to less memory transfer (Sec. 8.3). CPU libraries such as Alglib [7], GTSAM [12], and g2o [34] abstract the solver, requiring users to provide numeric routines for energy evaluation and, optionally, gradient calculation. All of these solvers create materialized Jacobians, and then use standard numerical linear algebra methods on these matrices to compute the Newton step. They cannot optimize the compilation of energy terms and solver code, unlike application-specific solvers, and require hand-written gradients to run fast. Similar to high-level solvers, Opt only requires a description of the energy, but it uses code transformations to generate application-specific matrix-free (or hybrid) GPU solvers automatically.

*Simulation DSLs.* Ebb [5] and Simit [32] are domain-specific languages that allow the user to express and abstract linear algebra compute operations over graphs (and in Ebb's case, arbitrary relations such as regular grids) on heterogeneous architectures. Ebb & Simit both focus on simulation, but could be used to write non-linear least squares solvers like those produced by Opt.

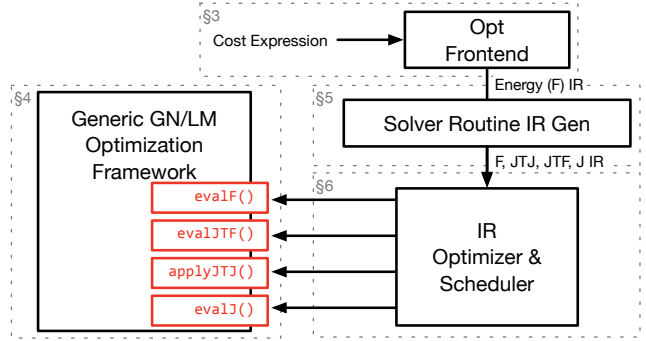

Fig. 2. An overview of the architecture of Opt, labeled with the sections where each part is described.

The user interface to Opt (specifying concise energy functions) is fundamentally a higher-level abstraction than the direct specification of arithmetic to execute in Simit and Ebb. To write an equivalent optimization in Simit or Ebb, a user must (1) write the solver algorithm from scratch; (2) manually derive efficient arithmetic code from the energy function (particularly tricky for fused Jacobian kernels); and (3) decide up front which parts of matrix multiplication are precomputed and cached vs. recomputed on the fly. Because Opt automates (1 & 2) and allows (3) to be specified post-hoc, users can iterate far more rapidly on their energy functions.

We prototyped a Gauss-Newton solver using Ebb for the Image Warping example, but found the solver generated by Opt was over 5x faster then the counterpart in Ebb. The implementation effort of the problem-specific solver in Ebb is similar to a hand-written CUDA implementation, thus significantly higher than specifying the energy in Opt. A future version of Opt could emit Simit or Ebb code, and that may be a practical solution to avoid maintaining multiple back-ends, but does not change the basic system design or address issues (1,2,3) laid out above.

*Differentiation Methods.* Matrix-free approaches require efficient derivative computation since the derivatives are evaluated in the inner iteration of the PCG loop. *Numeric differentiation*, which uses finite differences to estimate derivatives, is numerically unreliable and inefficient [26]. Instead, packages like Mathematica [65] allow users to compute *symbolic* derivatives using rewrite rules. Because they frequently represent math as trees, they do not handle common sub-expressions well, making them impractical for large expressions [26]. *Automatic-differentiation* is transformation on *programs* rather than symbols [22, 24]. They replace numbers in a program with "dual"-numbers that track a specific partial derivative using the chain rule. However, because the transform does not work on symbols, simplifications that result from the chain rule are not always applied. We use a hybrid *symbolic-automatic* approach similar to D⋆, which represents math symbolically but stores it as a directed acyclic graph (DAG) of operators to ensure scalability to large problems [26]. A symbolic representation of derivatives is important for Opt since solver routines use many derivative terms that share common expressions. This can not be addressed by auto-differentiation methods.

```
W,H = Dim("W",0), Dim("H",1)
X = Unknown("X",float,{W,H},0)
A = Array("A",float,{W,H},1)

w_fit,w_reg = .1,.9
Energy(w_fit*(X(0,0) - A(0,0)), --fitting
       w_reg*(X(0,0) - X(1,0)), --regularization
       w_reg*(X(0,0) - X(0,1)))
```

Fig. 3. Laplacian smoothing energy for a one-component image, implemented in Opt. Note that weights are the square root values, since the Energy function squares its inputs.

## 3 PROGRAMMING MODEL

An overview of Opt's architecture is given in Fig. 2. In this section, we describe our programming model to construct the problem specific energy functions. Sec. 4 describes our generic solver framework on GPUs, and describes the Gauss-Newton and Levenberg-Marquardt solvers we implemented using this framework. To operate matrix-free, this framework requires application-specific *solver routines* (evalF(),evalJ(),evalJTF(), applyJTJ()). Sec. 5 describes how our compiler generates these routines from the energy.

We introduce our programming model using the example of Laplacian smoothing of an image. A fitting term encourages a pixel $X$ to be close to its original value $A$:

$$E_{fit}(i,j) = [X(i,j) - A(i,j)]^2$$

A regularization term encourages neighboring pixels to be similar:

$$E_{reg}(i,j) = \sum_{(l,m) \in \mathcal{N}(i,j)} [X(i,j) - X(l,m)]^2$$

$$\text{where } \mathcal{N}(i,j) = \{(i+1,j),(i,j+1)\}$$

The energy is a weighted sum of both terms:

$$E_\Delta = \sum_{(i,j) \in \mathcal{I}} w_{fit} E_{fit}(i,j) + w_{reg} E_{reg}(i,j)$$

While this example is linear, Opt supports arbitrary non-linear energy expressions.

*Language.* Similar to shading languages such as OpenGL, Opt programs are composed of a "shader" file that describes the energy and a set of C APIs for running the problem. Fig. 3 expresses the Laplacian energy in Opt. Opt is embedded in the Lua programming language and operator overloading is used to create a symbolic representation of the energy. The first line specifies problem dimensions. Line 2 uses the function `Unknown` to declare the pixel array that represents the unknown $X$. `Array` is used to declare constant values such as the image $A$ that will be fixed during optimization. The last argument of these declarations is a numeric index that associates the array with actual data provided by the C API.

`Energy` adds residual expressions to the problem's energy. A key part of Opt's abstraction is that residuals are described at elements of images or graphs and are implicitly mapped over the entire domain. The term `w_fit*(X(0,0) - A(0,0))` defines an energy at each pixel that is the difference between the images. We support arrays and energies that include both vector and scalar terms. The `Energy` function implicitly squares the terms and sums them over the domain to enforce the linear least-squares model. Terms can also include a

```
void SolveLaplacian(int width, int height,
                    float* unknown, float* target) {
  OptState* state = Opt_NewState();
  // load the Opt DSL file containing the cost description
  OptProblem* problem = Opt_ProblemDefine(state,"laplacian.opt");
  // describe the dimensions of the instance of the problem
  uint32_t dims[] = { width, height };
  OptPlan* plan = Opt_ProblemPlan(state, problem, dims);
  // run the solver
  void* problem_data[] = { unknown_pixel_data, target_pixel_data };
  Opt_ProblemSolve(state, plan, problem_data, NULL);
}
```

Fig. 4. Opt API calls that use the Laplacian smoothing program.

```
N = Dim("N",0)
X = Unknown("X", float3,{N},0)
A = Array("A", float3,{N},1)
G = Graph("Edges", 2,
          "vertex0", {N}, 3,
          "vertex1", {N}, 4)

w_fit,w_reg = .1,.9
Energy(w_fit*(X(0) - A(0)), w_reg*(X(G.vertex0) - X(G.vertex1)))
```

Fig. 5. The Laplacian cost defined on the edges of a mesh instead of an image. The graph represents explicit connectivity.

statically-defined *stencil* of neighboring pixels. The regularization term `w_reg*(X(0,0) - X(1,0))` defines an energy that is the difference between a pixel and the pixel to its right. Our solver framework exploits this regularity to produce efficient code.

*API.* Applications interact with Opt programs using a C API. Fig. 4 shows an example using this API. To amortize the cost of preparing a problem used multiple times, we separate the compilation (`Opt_ProblemDefine`), memory allocation (`Opt_ProblemPlan`), and execution (`Opt_ProblemSolver`) of a problem into different API calls.

*Mesh-based problems.* Opt also includes primitives for defining energies on graphs to support meshes or other irregular structures. Fig. 5 shows an example that smooths a mesh rather than an image. The `Graph` function defines a set of hyper-edges that connect entries in the unknown together. In this example, each edge connects two entries `vertex0` and `vertex1`, but in general our edges allow an arbitrary number of entries to represent elements, such as three-element hyper edges to define triangles. Energies can be defined on these elements, as seen in the regularization term (line 10), which defines an energy on the edge between two vertices.

*Boundaries.* Defining energies on arrays of pixels requires handling boundaries. By default an entire energy term is considered to have zero energy if any of its accesses would be out of bounds, but we also provide the ability to have custom behavior by querying whether a pixel is valid (`InBounds`) and selecting a different expression if it is not (`Select`):

```
term = w_reg*(X(0,0) - X(1,0))
Energy(Select(InBounds(1,0),term,0))
```

Boundary handling is optimized later in the compilation process to ensure that it does not cause excessive overhead.

*Pre-computing shared expressions.* Energy functions for neighboring pixels can share expensive-to-compute expressions. For instance, our shape-from-shading example (Sec. 8) uses an expensive lighting calculation that is shared by neighboring pixels. We allow the user to turn these calculations into *computed arrays*, which behave like arrays when used in energy functions, but are defined as an expression of other arrays:

```
computed_lighting = ComputedArray(W,H,lighting_calculation(0,0))
```

Computed arrays can include computations using unknowns, and are recalculated as necessary during the optimization. Similar to scheduling annotations in Halide [48], they allow the user to balance recompute with locality at a high-level.

## 4 NON-LINEAR LEAST SQUARES OPTIMIZATION FRAMEWORK

Our optimization framework is a generalization of the design of application-specific GPU solvers based on the Gauss-Newton approach [30, 54, 55, 68, 72, 73]. However, our solver API abstracts away the specific algorithm details, allowing us to provide options for both Gauss-Newton and Levenberg-Marquardt approaches [6, 36, 39]. We first describe the approach our specific solvers use, and then show how we separate out the details of the application-specific energy from the solver being used.

In the context of non-linear least square problems, we consider the optimization objective $E : \mathbb{R}^N \rightarrow \mathbb{R}$, which is a sum of squares in the following canonical form:

$$E(\mathbf{x}) = \sum_{r=1}^{R} \left[ f_r(\mathbf{x}) \right]^2$$

The $R$ scalar residuals $f_r$ can be general linear or non-linear functions of the $N$ unknowns $\mathbf{x}$. The objective takes the traditional form used in the Gauss-Newton method:

$$E(\mathbf{x}) = \left\| \mathbf{F}(\mathbf{x}) \right\|_2^2, \ \mathbf{F}(\mathbf{x}) = [f_1(\mathbf{x}), \dots, f_R(\mathbf{x})]^T$$

The $R$-dimensional vector field $F : \mathbb{R}^N \rightarrow \mathbb{R}^R$ stacks all scalar *residuals* $f_r$. The minimizer $\mathbf{x}^*$ of $E$ is given as the solution of the following optimization problem:

$$\mathbf{x}^* = \underset{\mathbf{x}}{\operatorname{argmin}}\, E(\mathbf{x}) = \underset{\mathbf{x}}{\operatorname{argmin}} \left\| \mathbf{F}(\mathbf{x}) \right\|_2^2$$

It is solved based on a fixed-point iteration that incrementally computes a sequence of better solutions $\{\mathbf{x}_k\}_{k=1}^{K}$ given an initial estimate $\mathbf{x}_0$. Here, $K$ is the number of iterations; i.e., $\mathbf{x}^* \approx \mathbf{x}_K$. In every iteration step, a *linear* least squares problem is solved to find the best linear parameter update. The vector field $\mathbf{F}$ is first linearized using a first-order Taylor expansion around the last solution $\mathbf{x}_k$:

$$\mathbf{F}(\mathbf{x}_k + \delta_k) \approx \mathbf{F}(\mathbf{x}_k) + \mathbf{J}(\mathbf{x}_k)\delta_k$$

Here, $\mathbf{J}$ is the Jacobian matrix and contains the first-order partial derivatives of $\mathbf{F}$. By applying this approximation, the original non-linear least squares problem is reduced to a quadratic problem:

$$\delta_k^* = \underset{\delta_k}{\operatorname{argmin}} \left\| \mathbf{F}(\mathbf{x}_k) + \mathbf{J}(\mathbf{x}_k)\delta_k \right\|_2^2$$

After the optimal update $\delta_k^*$ has been computed, a new solution $\mathbf{x}_{k+1} = \mathbf{x}_k + \delta_k$ can be easily obtained. Since this problem is highly over-constrained and quadratic, the least squares minimizer is the solution of a linear system of equations. This system is obtained by setting the partial derivatives to zero, which results in the well known *normal equations*:

$$2 \cdot \mathbf{J}(\mathbf{x}_k)^T \mathbf{J}(\mathbf{x}_k)\delta_k^* = -2 \cdot \mathbf{J}(\mathbf{x}_k)^T \mathbf{F}(\mathbf{x}_k)$$

This process is iterated for $K$ steps to obtain an approximation to the optimal solution $\mathbf{x}^* \approx \mathbf{x}_K$.

The GN approach can be interpreted as a variant of Newton's method that only requires first-order derivatives and requires less computation. To this end, it uses a first-order Taylor approximation $2(\mathbf{J}^T \mathbf{J})$ instead of the real second-order Hessian $\mathbf{H}$.

LM additionally introduces a steering parameter $\lambda$ to switch between GN and Steepest Descent (**SD**). To this end, the normal equations are augmented with an additional diagonal term. This is similar to Tikhonov regularization and leads to:

$$2(\mathbf{J}(\mathbf{x}_k)^T \mathbf{J}(\mathbf{x}_k) + \lambda \operatorname{diag}\left(\mathbf{J}(\mathbf{x}_k)^T \mathbf{J}(\mathbf{x}_k)\right))\delta_k^* = -2\mathbf{J}(\mathbf{x}_k)^T \mathbf{F}(\mathbf{x}_k)$$

The inverse of $\lambda$ defines the radius of the *trust region*. LM guarantees convergence by shrinking the trust region radius and resolving the linear system when a proposed step fails to decrease cost. In the current LM solvers generated by Opt, we allow users to specify an initial trust region radius, minimum and maximum radii, and minimum and maximum values to clamp entries of the diagonal damping matrix. Our specific LM strategy is adapted from the Ceres solver [1]. In our comparisons, we use the same parameter configuration for both solvers.

### 4.1 Parallelizing the Optimization with PCG

The core of the GN/LM methods is the iterative solution of *linear least squares problems* for the computation of the optimal linear updates $\delta_k^*$. This boils down to the solution of a system of linear equations in each step, i.e., the *normal equations*. While it is possible to use direct solution strategies for linear systems, they are inherently sequential, while our goal is a fast parallel solution on a many-core GPU architecture with conceptually several thousand independent threads of execution. Consequently, we use a parallel preconditioned conjugate gradient (PCG) solver [63, 73], which is fully parallelizable on modern graphics cards.

The PCG algorithm and our strategy to distribute the computations across GPU kernels is visualized in Fig. 6. We run a *PCGInit* kernel (one time initialization) and three *PCGStep* kernels (inner PCG loop). Before the PCG solve commences, we initialize the unknowns $\delta_0$ to zero. For preconditioning, we employ the Jacobi preconditioner, which scales the residuals with the inverse diagonal of $\mathbf{J}^T \mathbf{J}$. Jacobi preconditioning is especially efficient if the system matrix is diagonally dominant, which is true for many problems; for instance, the Laplacian operator and most of its variations are diagonally dominant. When the matrix is not diagonally dominant, we fall back to a standard conjugate gradient descent by user selection. More general preconditioners could be provided as a parameter at code generation time but are not a focus of this paper. A detailed overview of different preconditioning approaches in parallel solvers is given in [23], and matrix-free preconditioners are proposed by [4, 69]. We also default to single-precision floating point numbers throughout, which matches the approach of the recent application-specific

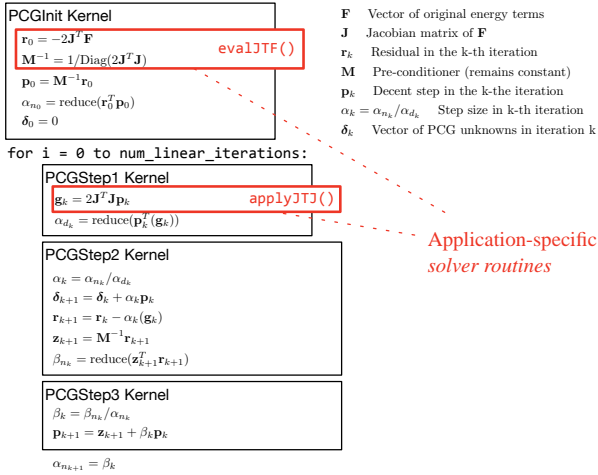

Fig. 6. Generic GPU architecture for Gauss-Newton and Levenberg-Marquardt solvers whose linearized iteration steps are solved in parallel using the preconditioned conjugate gradient (PCG) method.

solvers. We believe this strategy is a good compromise between computational effort and efficiency.

*Stencil-based Array Access.* Our techniques for parallelizing work are different for array and graph residuals. For arrays, we group the computation required for each element in the *unknown* domain onto one GPU thread. For a matrix product such as $-2\mathbf{J}^T\mathbf{F}$, each row of the output is generated by the thread associated with the unknown. If the unknown is a vector (e.g., RGB pixel), all channels are handled by one thread since these values will frequently share sub-expressions.

The computations in a GPU thread can work *matrix-free*. For instance, if they conceptually require a particular partial derivative from matrix $\mathbf{J}$, they can compute it from the original problem state, which includes the unknowns and any supplementary arrays. Matrices such as $\mathbf{J}$, which are conceptually larger than the problem state, do not need to be written to memory, which minimizes memory accesses. Section 5 describes how we automatically generate these computations from our stencil- and graph-based energy specification.

*Graph-based Array Access.* For graph-based domains, such as 3D meshes, the connectivity is explicitly encoded in a user-provided data structure. Users specify the mapping from graph edges to vertices. Residuals are defined on graph (hyper-) edges and access unknowns on vertices. To make it easy for the user to change the graph over time, we do not require a reverse mapping from unknowns to residuals for graphs. Kernels that use the residuals (`PCGInit` and `PCGStep1`) assign one edge in the graph to one GPU thread. Since the output vectors have the same dimension as the unknowns, we have to *scatter* the terms in the residual evaluations into these values. All threads involving partial sums for a given variable then scatter into the corresponding parts of variables using a floating-point atomic addition.

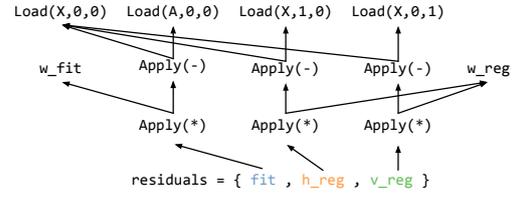

Fig. 7. The Laplacian example represented in our IR.

## 4.2 Modularizing the Solver

A key contribution of our approach is the modularization of the application-specific components of GPU Gauss-Newton or Levenberg-Marquardt solvers into compartmentalized *solver routines*. The first routine, `evalF()`, simply generates the application specific energy for each residual. It only runs outside of the main loop to report progress.

*evalJTF.* The second routine appears in the `PCGInit` kernel and is shown in red in Fig. 6. Here, the initial descent direction $\mathbf{p}_0$ is computed using the application-specific `evalJTF()` routine, which is generated by our compiler. It computes a matrix-free version of $-2\mathbf{J}^T\mathbf{F}$. `evalJTF()` is also responsible for computing the preconditioner $\mathbf{M}$, which is simply the dot product of a row of $\mathbf{J}^T$ with itself. For arrays, a thread computes the rows of an unknown associated with one element of the unknown. For graphs, each thread only computes the parts of the dot product between $\mathbf{J}^T$ and $\mathbf{F}$ which belong to the handled residual.

*applyJTJ.* The third routine, `applyJTJ()`, is part of the inner PCG iteration. It computes the multiplication of $2\mathbf{J}^T\mathbf{J}$ with the current descent direction $\mathbf{p}_k$, and incorporates the steering factor $\lambda$ when using Levenberg-Marquardt. Handling arrays and graphs is similar to `evalJTF()`. It tends to use more values since it needs to compute entries from both $\mathbf{J}$ and $\mathbf{J}^T$. For many problems this routine is the most expensive step, so it has to be optimized well.

*evalJ.* While `evalJTF` and `applyJTJ` are used in matrix-free code, in some cases materialized matrices are faster. In these cases, our solvers can use the `evalJ` routine which calculates individual entries in $\mathbf{J}$ that the solver can materialize in memory.

## 5 GENERATING SOLVER ROUTINES

A key idea of Opt is that we can exploit the regularity of stencil- and graph-based energies to automatically generate application-specific solver routines. We represent the mathematical form of the energy as a DAG of operators which we refer to as our *intermediate representation* (IR). We transform the IR to create new IR expressions needed for `evalJTF()`, `applyJTJ()`, and `evalJ()`. This process requires partial derivatives of the energy. We then optimize this IR and generate code that calculates it.

### 5.1 Intermediate Representation

Since the Opt language is embedded in Lua, we generate the IR by running the Lua program which uses overloaded operators to build the graph. Fig. 7 shows the IR that results from the Laplacian

```
-- generates the derivative of expression with respect to variable
function derivative(expression, variable)
    if a cached version of this partial derivative exists then
        return the cached version
    elseif expression == variable then
        return 1
    end
    result = 0
    for i = 0, the number of arguments used by expression do
        result += derivative(argument[i],variable)*partial[i]
    end
    cache and return result
end
```

Fig. 8. Pseudocode of the OnePass algorithm for generating derivatives. `partial[i]` is the partial derivative of the particular operator (e.g., $\star$) with respect to the argument i, which is defined for each operator.

example. Roots of the IR are residuals that we want to compute. Leaves are constants (e.g., w_fit), input data (e.g the known image A(0,0)), and the the unknown image (e.g., X(0,0)). We de-duplicate the graph as it is built, ensuring common-subexpressions are eliminated. We scalarize vectors from our frontend in the IR to improve the simplification of expressions that become zeros during differentiation.

### 5.2 Differentiating IR
Since we do not always store the Jacobian $\mathbf{J}$ in memory, we need to generate residuals on-the-fly. The approach we use for differentiation is similar to Guenter's D$\star$ [26]. It symbolically generates new IR that represents a partial derivative of an existing IR node. Unlike traditional symbolic differentiation (e.g., Mathematica), differentiation is done on a graph where terms can share common sub-expressions. In our implementation we use OnePass, a simplification of D$\star$ that can achieve good results by doing the symbolic equivalent of forward auto-differentiation [27]. Pseudocode for the algorithm is given in Fig. 8. It works by memoizing a result for each partial derivative and generates a new derivative of an expression by propagating derivatives from its arguments via the chain rule.

### 5.3 Generating IR for Matrix Products
The IR for `evalF()` is simply the input energy IR. We generate IR for `evalJTF()`, `applyJTJ()`, and `evalJ()` as transformations of this input IR. The first two terms are conceptually derived from matrix-matrix or matrix-vector multiplications of the Jacobian. Since we compute these values matrix-free, we must generate the IR that will calculate the output given our specific problem. Each term has two versions: one for handling stencil-based and one for graph-based residuals.

#### 5.3.1 Stencil Residuals.
Our solver calls `applyJTJ()` to calculate a single entry of $\mathbf{g}$, where $\mathbf{g} = 2\mathbf{J}^T\mathbf{J}\mathbf{p}$ per thread. We need to determine which values from $\mathbf{J}$ are required and create IR that calculates them. The non-zero entries in $\mathbf{J}$ are determined by the *stencil* of a particular problem. Fig. 9 illustrates the process of discovering the non-zeros. In the Laplacian case, the partials used in these expressions are actually constants because it is a linear system. However, Opt supports the generic non-linear case, where the partials will be functions of the unknown.

The pseudocode to generate $\mathbf{J}^T\mathbf{J}$ for stencils is shown in Fig. 10. It first finds the residuals that use unknown $\mathbf{x}_{0,0}$ because they correspond to the non-zeros of $\mathbf{J}^T$. Some of these residuals are not actually *defined* at pixel (0, 0), but use $\mathbf{x}_{0,0}$ from neighboring pixels. To find them, we exploit the fact that stencils are *invertible*. For each residual *template* in the energy, we examine each place it uses an unknown $\mathbf{x}_{i,j}$. We then *shift* that residual on the pixel grid, taking each place it loads a stencil value and changing its offset by $(-i, -j)$, which generates a residual in the grid that uses $\mathbf{x}_{0,0}$. We find all the residuals using $\mathbf{x}_{0,0}$ by repeating the process for each use of an unknown in the template. While we only allow constant stencil offsets, in principle this approach will work for any neighborhood function which is invertible.

For each discovered residual, we need the other unknowns it uses which are found by examining the IR symbolically. We then generate the expressions for the part of the matrix-vector products that calculate $\mathbf{g}_{0,0}$. In this code, we symbolically compute the partial derivatives that are the entries of $\mathbf{J}$.

#### 5.3.2 Graph Residuals.
For graphs, residuals are defined on hyper-edges rather than on the domain of the unknown and our solver routines are mapped over residuals directly so we do not need an inverse mapping from unknown to residual. Instead one thread computes the part of an output that relates to the residual. Pseudocode to generate `applyJTJ()` for graphs is given in Fig. 11. At each residual, it generates one row of $\mathbf{J}\mathbf{p}$, and then performs the part of the multiplication for the rows of $\mathbf{g}$ that include partials for that residual. The output of this routine is a list of IR nodes that are atomically added into entries of $\mathbf{g}$.

#### 5.3.3 Variants.
The approach to generate `evalJTF()` and `evalJ()` is similar to that of `applyJTJ()`. The routine `create_jtf()` is used to generate the expression $r = -2\mathbf{J}^T\mathbf{F}$. Each row of $\mathbf{J}^T$ can be obtained using the same approach previously described. The partials in this row are then multiplied directly with their corresponding residual term in $\mathbf{F}$. Similarly, the routine `create_j()` simply produces all the non-zero derivatives for a particular residual that can be stored by the solver for use in materialized approaches.

In LM we additionally need the term $\lambda \operatorname{diag}(\mathbf{J}^T\mathbf{J})$, which is inserted into the `applyJTJ()` routine when needed.

## 6 OPTIMIZING GENERATED SOLVER ROUTINES
We need to translate the IR for `evalF()`, `evalJTF()`, `applyJTJ()`, `evalJ()` into efficient GPU functions. We simplify the IR based on polynomial simplification rules, optimize the handling of boundary condition statements, and schedule the IR which generates GPU code.

*Polynomial Simplification.* Taking the derivative of the IR tends to introduce more complicated IR. In particular, the application of the multi-variable chain rules introduces statements of the form $d_1 * p_1 + d_2 * p_2 + \ldots$ for each argument of an operator. Often some partials are zero, and terms in the sum can be grouped together. We take the approach of other libraries like SymPy [52] and represent primitive math operations as polynomials. In particular, additions and multiplications are represented as *n*-ary operators rather than binary, and we include a `pow` operator that raises an expression to a constant $a^c$. Where possible, primitives are represented in terms of these operators. For instance $a/b$ is represented as $ab^{-1}$ and $a - b$ as $a + -1 * b$. Polynomial representation makes it easier to find opportunities for optimization such as constant propagation when the optimization first requires re-associating, commuting, or factoring expressions.

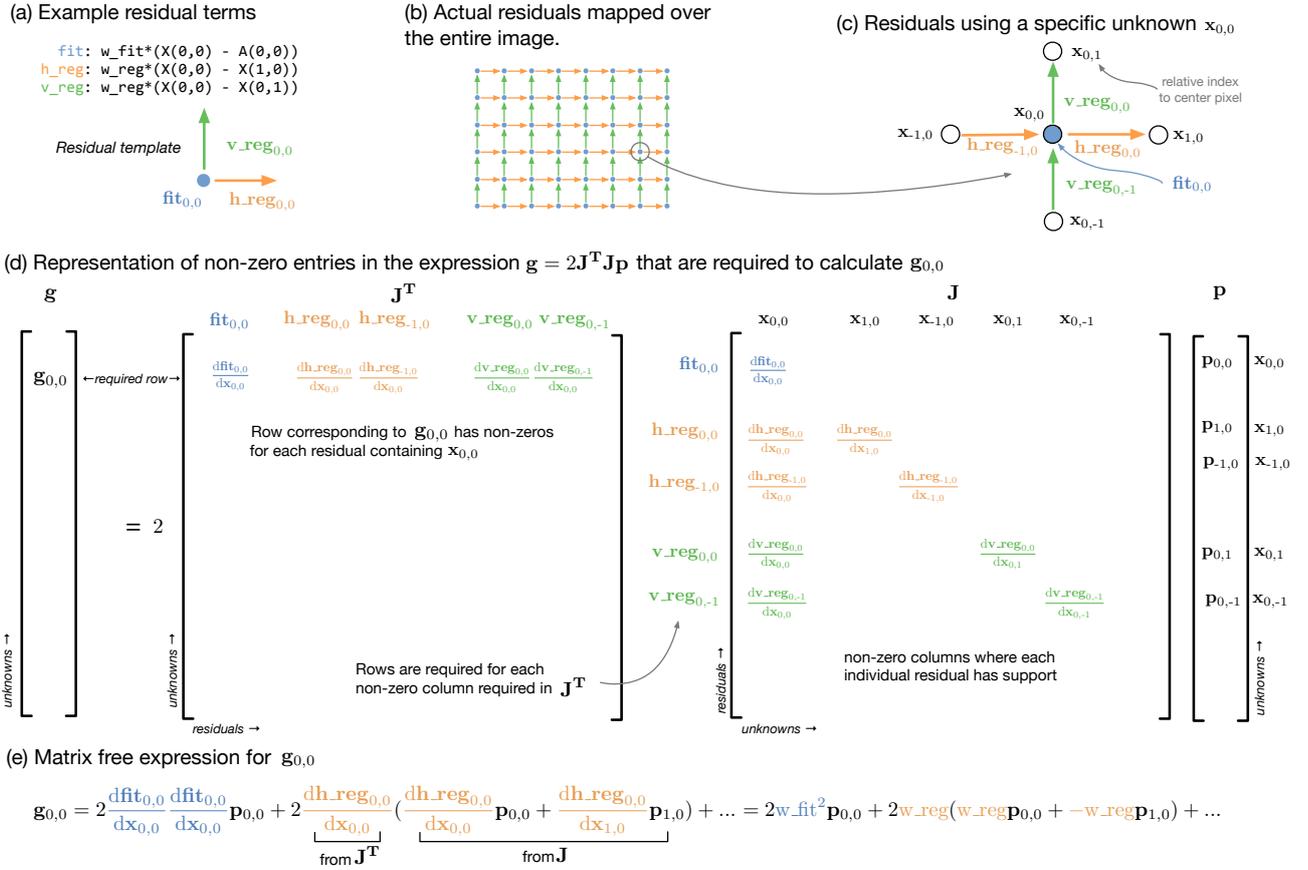

Fig. 9. The process our compiler uses for generating `applyJTJ()` at a high-level. (a) The input to this transformation is a list of individual residuals defined in the IR (`fit`, `h_reg`, `v_reg`) that form a *template*. (b) The residual template is repeated over the image to generate the actual energy function. (c) The compiler considers a specific unknown $\mathbf{x}_{0,0}$, here shown with the residuals that refer to it, which are computed by the compiler. Unknowns and residuals are named relative to this pixel (e.g., `h_reg_1,0` is the horizontal residual from the pixel to the left). (d) The compiler then symbolically forms the result of $\mathbf{g} = 2\mathbf{J}^{\mathbf{T}}\mathbf{Jp}$, here shown with the components needed to generate $\mathbf{g}_{0,0}$. The row of $\mathbf{J}^T$ corresponding to unknown $\mathbf{x}_{0,0}$ is needed. It has one non-zero for each residual in (c). This row will be multiplied against $\mathbf{Jp}$. The only rows of $\mathbf{Jp}$ needed correspond to the residuals appearing in $\mathbf{J}^T$ since other rows will be multiplied by 0. A row of $\mathbf{Jp}$ is calculated by multiplying non-zero entries in a row of $\mathbf{J}$, which occur each time a residual uses an unknown, against the corresponding row of $\mathbf{p}$. (e) Finally, the compiler forms a matrix-free version of the expression for $\mathbf{g}_{0,0}$ implied by the matrix multiplications, calculating each partial using one-pass differentiation.

Importantly, the polynomial representation also gives our scheduler freedom to reorder long sums and products to achieve other goals, such as grouping terms with the same boundary statement into a single if-statement or minimizing register pressure.

During construction we optimize non-polynomial terms using constant propagation and applying algebraic identities. Before lowering into code, we also apply a factoring pass that applies a greedy multi-variate version of Horner's scheme [10] to pull common factors out of large sums.

*Bounds Optimization.* Boundary conditions introduce another source of inefficiency. Opt uses `InBounds` and `Select` to create boundary conditions and masks. Translating these expressions to code can introduce inefficiency in two ways. First, it is possible for the same bound to be checked multiple times. This frequently occurs in `applyJTJ()` when two partials are multiplied together since both partials often contain the same bound. Redundant checks also occur when reading from arrays since Opt must always check array bounds to avoid crashes. This check is often redundant with a `Select` already in the energy. Secondly, without optimization, `Select` statements need to execute both the true and false expressions. For many cases, this means that large parts of the IR, including expensive reads from global memory, do not actually need to be calculated but are performed anyway.

```
function create_jtj(residual_templates,X,P)
    P_hat = 0
    residuals = residuals_including_x00(residual_templates)
    foreach residual do
        dr_dx00 = differentiate(residual,X(0,0))
        foreach unknown u used by residual do
            dr_du = differentiate(residual,u)
            P_hat += dr_dx00*dr_dx*P(u.offset_i,u.offset_j)
        end
    end
    return 2*P_hat
end
function residuals_including_x00(residual_templates)
    residuals = {}
    foreach residual_template do
        foreach unknown x appearing in residual_template do
            -- shift the template such that x is centered (i.e. it is x00)
            R = shift_exp(residual_template,-x.offset_i,-x.offset_j)
            table.insert(residuals,R)
        end
    end
    return residuals
end
function shift_exp(exp, shift_i, shift_j)
    replace each access of any image at (x,y) in exp
    with an access at (x + shift_i,j + shift_j)
end
```

Fig. 10. Pseudocode of the compiler transformation that Opt uses to generate $\mathbf{J}^T\mathbf{J}$ from residual templates.

```
function create_jtj_graph(graph_residuals)
    foreach graph_residual do
        Jp = 0
        -- handle Jp multiply against this residual
        foreach unknown u appearing in graph_residual do
            dr_du = differentiate(graph_residual,u)
            Jp += dr_du*P(u.index)
        end
        -- handle partial sums for Jt*Jp
        foreach unknown u appearing in graph_residual do
            dr_du = differentiate(graph_residual,u)
            insert atomic scatter:
                P_hat(u.index) += 2*dr_du*Jp
        end
    end
    return set of atomic scatters
end
```

Fig. 11. Pseudocode of the compiler transformation that Opt uses to generate $\mathbf{J}^T\mathbf{J}$ for graph residual terms.

The common approach of generating two versions of code, one for the boundary region and a bounds-free one for the interior, is less effective on GPUs because they group threads into wide vector lanes of 32 elements, which increases the size of the boundary by the vector width. For smaller sized problems, large portions of the image fall in the boundary region.

Instead, we address these two sources of inefficiency directly. We address the redundant bounds checks by augmenting our polynomial simplification routines to handle bounds as well. We represent bounds internally as polynomials containing boolean values $b$ that are either $0$ or $1$. A `Select(b,e_0,e_1)` is then represented as $b*e\_0 + \sim b*e\_1$. We simplify booleans raised to a power $b^e$ to $b$. This representation allows polynomial simplification rules to remove redundant bounds through factoring. We favor booleans over other values during factoring to ensure this simplification occurs.

We address excessive computation and memory use due to bounds by determining when values in the IR need to be calculated. We associate a boolean *condition* with each IR node that conservatively bounds when it is used. These conditions are generated by `Select` statements and propagated to their arguments. To improve the effectiveness of this approach, we split large sums into individual reductions that update a summation variable. Each reduction can then be assigned a different condition. When we actually schedule code, we will only execute the code if its condition is true.

*Scheduling and Code Generation.* We translate optimized IR into actual GPU code by *scheduling* the order in which the code executes the IR. Our scheduler uses a greedy approach that is aware of our boundary optimizations. It starts with the instructions that generate the output values and schedules backwards, maintaining a list of nodes that are *ready* to be scheduled according to their dependencies. It iteratively chooses an instruction from the ready list that has the lowest *cost*, schedules it, and updates the list. Our cost function first prioritizes scheduling an instruction with the same *condition* as the previous instruction, grouping expressions that have the same bounds together into a single if-statement. It then prioritizes choices that greedily minimize the set of live variables at that point in the program, which can provide a small benefit for large expressions. We also prioritize the instruction that has been ready the longest, which also helps reduce the required registers [57].

We translate the scheduled instructions into GPU code using Terra [16]. Terra is a multi-stage programming language with metaprogramming features that allow it to generate high-performance code dynamically. We use its GPU backend to produce CUDA code for the solver routines. To improve the performance, we automatically generate code to bind and load input data from GPU **textures**. In addition to having better caching behavior, textures also can perform the bounds check for loads automatically. Finally, the solver routines are inlined into the generic solver framework presented earlier. Because this code is compiled together, there is no overhead when invoking solver routines.

## 7 METAPROGRAMMING FLEXIBILITY

Our architecture separates the specification of the energy using the Opt language from the specification of the LM/GN solvers, which interact with the energy only through the abstract solver routines generated by the compiler. This design facilitates various forms of experimentation to produce fast and effective solvers. These choices can be made quickly by changing flags in Opt.

*LM vs. GN.* For matrix-free code, it can take significant effort to add LM extensions to a custom GN solver just to check if they are needed. The application of $\mathbf{J}^T\mathbf{J}$ specifically requires additional energy-specific code that can involve potentially complicated derivatives. In Opt, a single flag enables LM while leveraging the GN solver routine generators that are combined with the LM-specific extensions. This allows for the speed of GN when possible, but the convergence guarantees of LM when necessary.

*Matrix-free vs. Materialized.* A key insight of previous hand-written GPU methods [73] adapted by our framework is that it is sometimes more efficient to compute $\mathbf{J}$ in-place rather than store $\mathbf{J}$ or $\mathbf{J}^T\mathbf{J}$ as a sparse matrix. This approach can be faster for two reasons. First, the locations of the non-zero entries in the matrix are implicitly represented by the problem domain (either an image or a graph), and are not loaded explicitly. Second, entries in the matrix can often be recomputed using less total memory bandwidth than loading the full $\mathbf{J}$ matrix. In the extreme case, such as Poisson Image Editing described in the next section, the matrix is constant and the non-zeros can be folded directly into the code. However, this does not apply for compute-intense problems such as fully-differentiated Cotangent-weighted Laplacian Smoothing, described in the appendix, where the compute cost dominates bandwidth.

Opt addresses this trade off by allowing the (gradual) choice between matrix-free and fully-materialized operations. Opt can use either a matrix-free or materialized approach. The matrix-free approach uses `applyJTJ()`, while the materialized approach uses `evalJ()` to materialize $\mathbf{J}$ to GPU memory.

In our current implementation, we then use cuSPARSE, a high-performance GPU sparse matrix library, inside our materialized PCG solver [46]. This pathway can be extended to work with any GPU sparse solver.

We can also represent hybrid approaches where expensive intermediates are materialized using `ComputedArray` annotations in the energy. The remaining parts of the computation are still computed on demand. This middle ground is sometimes more efficient and is easy to investigate using annotations; see our Shape from Shading example in Fig. 17.

*Numerical Precision.* Opt also allows users to switch between float and double precision depending on the needs of the application and the capability of the GPU compute platform. Although most our graphics example problems are well-conditioned enough for floating point precision, one might want to trade speed for more stability for ill-conditioned problems; for a detailed numeric evaluation on a standard optimization benchmark, we refer to the appendix.

*Variants of Standard LM/GN.* Many other kinds of solvers are also just variants of LM/GN that can fit into the Opt model. For instance, $\ell_p$ problems of the form

$$E(\mathbf{x}) = \sum_{r=1}^{R} \left[ f_r(\mathbf{x}) \right]^p$$

which solve for norms other than $L_2$ can be computed using Iteratively re-weighted least squares (**IRLS**), which iteratively solves the least squares problem

$$E(\mathbf{x}) = \sum_{r=1}^{R} w_i \left[ f_r(\mathbf{x}) \right]^2$$

where $\left[ f_r(\mathbf{x}) \right]^2$ is a normal least squares function and $w_i = \left[ f_r(\mathbf{x}) \right]^{p-2}$ is fixed for the current iteration. Opt realizes these iterative solves by specifying the $w_i$ computation as fixed rather than part of the unknown; thus, no derivatives are computed. The example of Intrinsic Images, described in detail in the appendix, uses this type of solver to enforce sparsity on the solution.

Robust kernels are another common approach for non-linear least squares optimization problems in computer vision. Here, auxiliary variables are introduced in order to determine the relevance of a data term in part of the optimization formulation. The Robust Mesh Deformation example in the appendix shows how this approach can be naturally expressed in Opt.

For some problems, such as Dense Optical Flow, described in the appendix, unknowns are used to sample values from constant images. We support this pattern using a *sampled image* operator, which can be accessed with arbitrary $(u, v)$ coordinates. When these coordinates are dependent on the unknown image, the user provides the directional derivatives of the sampled image as other input images, which will be used to lookup the partials for the operator in the symbolic differentiation.

*Domains.* Opt is able to exploit the implicit structure and connectivity of general n-dimensional arrays. In addition to images, optimizations are often performed on volumetric grids (e.g., [30, 72] or time-space (e.g., [62]) domains, all of which are subsets of n-D arrays and fall within the scope of Opt. Volumetric Mesh Deformation, as described in the appendix, is an example of solving for unknowns on a 3-D array.

Opt efficiently handles large numbers of unknowns at each location in a regular array. For example, the Embedded Deformation example, described in the appendix, uses 12 unknowns per vertex.

Opt also efficiently handles explicit structure, provided in the form of general graphs. These domains include manifold meshes and general non-manifolds. For instance, non-rigid mesh deformation approaches (e.g., [50, 51]) fall into this category, as well as widely-used global bundle adjustment methods [2, 49, 58]. Cotangent Laplacian Smoothing, described in the appendix, provides a graph connecting the wedge of triangles at each edge together using graph hyper-edges.

Our abstraction also allows the energies on *mixed* domains. For example, an objective may contain dense regularization terms affecting every pixel of an image and a sparse set of correspondences from a fitting term. Here, the regularization energy is implicitly encoded in a 2D image domain, and the data term may be provided by a sparse graph structure.

On all of these domains, Opt provides automatic derivation of objective terms, and generates GPU solvers specifically optimized for a given energy function at compile time.

*Multi-pass Optimization.* In many scenarios, solving a single optimization is not enough, but instead requires multiple passes of different non-linear solves. Often, hierarchal, coarse-to-fine solves are used to achieve better convergence, or sometimes problem-specific flip-flop iteration can be applied (e.g., ARAP flip-flop by Sorkine and Alexa [50]). Another common case are dynamic changes in the structure of the optimization problem. For instance, fitting a mesh to point-cloud data in a non-rigid fashion is typically achieved by searching for correspondences between optimization passes (e.g., non-rigid iterative closest point) [38, 73]. Changes to the correspondences also change the structure of the sparse fitting terms.

In all of these examples, custom code is required at specific stages during optimization. To support this code in Opt, we take an approach similar to multi-pass rendering in OpenGL. Between iterations of the Opt solver or between entire solves, users can perform arbitrary modifications to the underlying problem state in C/C++. Optimization weights can be changed (e.g., for parameter relaxation), underlying data structures may be dynamically updated (e.g., correspondence search or feature match pruning in bundle adjustment problems), or hierarchical and flip-flop strategies can be applied using multiple-passes. This approach allows Opt to support a wide range of solver approaches, while still providing an efficient optimization backend for their inner kernels.

| Energy | Length in Opt |
|---|---|
| **ARAP Image Warping**   Interactively edit 2D images by warping them using an as-rigid-as-possible warping energy. | 21 lines vs. 280 custom |
| **ARAP Mesh Deformation**   Deform a mesh using an as-rigid-as-possible warping energy. | 18 lines vs. 200 custom |
| **Shape From Shading**   Refine depth data from RGB-D scanners using a detailed color image and an estimate of lighting based on spherical harmonics. | 96 lines vs. 445 custom |
| **Poisson Image Editing**   Splice a source image into a target image without introducing seams. | 13 lines vs. 67 custom |
| **3D LARAP Mesh Deformation**   Warp a mesh using an underlying 3D volumetric grid. | 21 lines |
| **Embedded Deformation**   Perform mesh deformation by solving a full affine transformation per vertex. | 34 lines |
| **Cotangent Mesh Smoothing**   Smooth a mesh while preserving the areas of triangles adjacent to each edge. | 32 lines |
| **Optical Flow**   Compute the apparent motion of objects between frames of video at the pixel level. | 20 lines |
| **Robust Mesh Deformation**   A version of ARAP Mesh Deformation that adds a robust kernel. | 27 lines |
| **Intrinsic Image Decomposition**   Separate an image into its reflectance and shading components. | 32 lines |

Fig. 12. Example applications written in Opt used in our evaluation. As a proxy for simplicity of implementation, lines of code for the energy in Opt are listed on the right, along with lines of code for the energy-specific code required by handwritten custom solvers when available. Both numbers do not include CPU code for data marshaling and setup.

## 8   EVALUATION

To evaluate Opt, we implemented several optimization problems in the language which are summarized in Fig. 12, and described in more detail in the appendix. These include variants of image, volume, and mesh-based problems from the graphics and vision literature. We evaluate overall performance by comparing Opt to four state-of-the-art application-specific matrix-free solvers optimized for GPUs and to five solvers using the high-level Ceres library [1]. We further evaluate the benefits and tradeoffs of Opt's ability to generate matrix-free, fully-materialized, or intermediate solver variants. We also show the efficiency of our automatically-generated solver routines (e.g., `applyJTJ`) by comparing them to hand-optimized equivalents. Finally, we implement five other problems which demonstrate the generality and expressiveness of Opt, referenced in Sec. 7, and described in detail in the appendix. The Opt code used for the energies of each example is also provided in the appendix.

Results are reported as throughput on entire solve steps using a GeForce 1080 GTX, and for CPU results, an Intel Core i7-6700K CPU 4.00 GHz.

## 8.1   Comparison with Custom Solvers

We compare solvers generated by Opt to existing state-of-the-art CUDA-based application-specific matrix-free solvers optimized for GPUs for ARAP Image Warping, Shape From Shading, ARAP Mesh Deformation, and Poisson Image Editing. Each of the original solvers took months to write, debug, and optimize in CUDA. As a concrete example, debugging the hand-written matrix-free application routine in the custom ARAP image warping originally took weeks due to the complicated cross terms that create dependencies between offsets of one pixel and the angles at a neighbor. In these comparisons, we select the Gauss-Newton backend of Opt to match the algorithmic design in the hand-written reference implementations.

The Opt solvers are both significantly easier to write and faster than the handwritten application-specific solvers. In Opt, the energies for each problem could be written in tens of lines of code (Fig 12). Furthermore, Opt outperforms the handwritten solvers for all these example problems by 10-75%, see Fig. 13.

Opt can outperform custom solvers because all Opt solvers benefit from optimizations made to the system. The Opt solver for ARAP Mesh Deformation runs 55% faster than custom code due to our reduction-based approach for calculating residuals. In the original solver, the authors only tried the simpler approach of using one pass to compute $\mathbf{t} = (\mathbf{Jp})$ and a second for $\mathbf{J}^T\mathbf{t}$. Opt's high-level model allowed us to experiment with different approaches more easily during development. In Shape From Shading, the Opt solver runs more than 30% faster than the handwritten CUDA solver. Some of this improvement is due to using texture objects to represent the images, which is an optimization that the original authors did not have time to implement. The ARAP Warping solver generated by Opt runs about 75% faster (likely due to better bounds handling) than the handwritten CUDA solver we compare against.

Since Poisson Image Editing is a linear problem, we also compare against a custom Cholesky solver with pre-ordering using Eigen [25], a high-performance linear-algebra library for CPUs. The Gauss-Newton method handles *linear* least-squares problems in a unified way that does not require algorithmic changes. When all residuals are linear functions of the unknowns, $\mathbf{J}$ just becomes a constant matrix independent of $\mathbf{x}$. All second order derivatives are zero, which implies that the Gauss-Newton approximation is exact and the optimum can be reached after a single non-linear iteration. The entire Opt solve was 50 times faster than Eigen's matrix solve (not including its matrix setup time), due to Opt's ability to implicitly represent the connectivity of the matrix.

## 8.2   Comparison with General Purpose Solver

We also compare Opt against the high-level Ceres library, which is also able to generate a solver using only an energy specification but does not support GPU or matrix-free execution. The solvers generated by Opt are 1–3 orders of magnitude faster than Ceres on our example problems (Fig 13). For accurate comparison, we setup both Opt and Ceres to use the same LM configuration, and plot their convergence over time in Fig 14. To get the fastest results for the internal linear system, we configure Ceres to use its parallel PCG solver for Image Warping and Shape From Shading, and Cholesky factorization for Mesh Deformation.

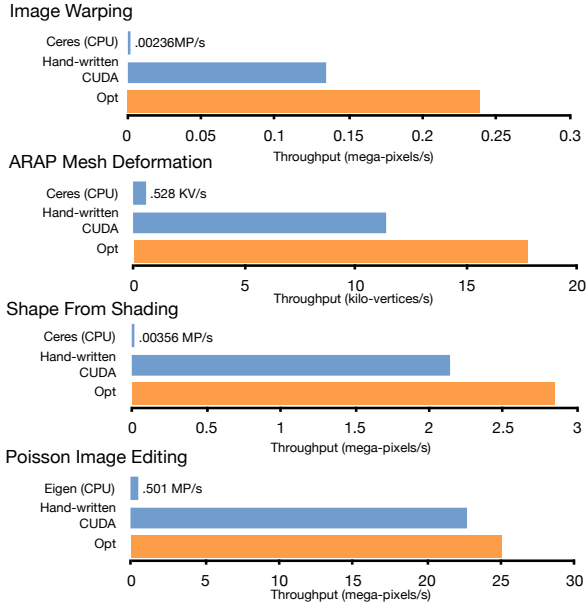

Fig. 13. The solvers generated by Opt perform better than application-specific GPU solvers, despite requiring significantly less effort to implement. Additionally, they outperform Ceres implementations by up to three orders of magnitude, despite requiring similar implementation effort.

One reason Opt is faster than Ceres is because Opt can represent the connectivity of problems on image and $n$-D array domains implicitly through stencil relations, while Ceres requires the user to specify energies using a graph formulation. The performance difference for Mesh Deformation is less dramatic than image-based examples because in this case Opt needs to load the connectivity of the problem from the graph data structure. However, Opt still benefits from repetitive stencil terms that are embedded in the generated code, as well as the massive parallelization of the GPU-based solver and on-the-fly computations. The Opt generated solver runs over 720 times faster on Shape From Shading, a relatively complex problem, due in part to a smart materialization strategy enabled by Opt, see Sec. 8.3. The performance benefits of Opt become more pronounced as problem size increases (Fig 15).

### 8.3 Matrix-free vs. Materialized Solvers

As mentioned in Sec. 7, a powerful property of Opt is the ability to use matrix-free representations or hybrid representations while still supporting fully materialized solvers. We show the difference between matrix-free and materialized approaches in Fig 16. Here, Opt uses cuSPARSE for the inner multiply of the PCG solver [46]. Note that cuSPARSE only provides the functionality for linear algebra and by itself it cannot tackle non-linear least squares problems due to a lack of auto- or symbolic differentiation.

Except for the highly non-linear Cotangent-weighted Laplacian Smoothing problem, all examples perform between 1.16 and 3 times faster using matrix-free approaches.

Opt also allows for intermediate materialization strategies, which allows users to choose which terms to precompute at the beginning of each non-linear iteration, using the ComputedArray construct described in Sec. 7.

We show the performance of the linear iterations for different materialization strategies on Shape From Shading in Fig. 17. The linear iterations are most efficient when we materialize the (compute-intensive) lighting term and its gradient, but recompute the rest of the Jacobian every linear iteration in a matrix-free approach.

### 8.4 Implicit vs. Explicit Connectivity

The examples throughout the paper demonstrate that Opt can handle both implicit connectivity on regular grids and explicit connectivity as specified by hypergraphs. It is difficult to quantify the performance improvement due to using an implicit representation of connectivity in general, but we provide a comparison between the two approaches on the Image Warping example in Fig. 18. We compare the performance of Opt using a Gauss-Newton solver over the standard regular grid representation of an image, and Opt using a Gauss-Newton solver over an explicit graph representation of the image. For very small image sizes the performance difference is minimal, but as image size increases the explicit approach takes about twice the amount of time to complete. The implicit approach saves both memory and bandwidth.

### 8.5 Evaluation of Generated Solver Routines

Our approach relies on the symbolic translations of energy functions into efficient solver routines using the optimizations described in Sec. 6. Compared to hand-written code, this code is much easier to write and maintain, but inefficient translations could make it too slow. To show the effectiveness of our symbolic translations and optimization, we compare our generated solver routines to hand-written versions that were taken from the pre-existing CUDA code and slotted into our solver.

Fig. 19 shows the results of our optimizations compared to the hand-written versions of solver routines ported from the CUDA examples and modified to use texture loads. The baseline (labeled "none" in the figure) roughly simulates how an auto-differentiation approach based on dual numbers would perform.

Our optimizations increase performance up to 8x in the case of Shape From Shading, and are necessary for Opt to perform at or above the speed of hand-written code. Performing polynomial simplifications improves the results of all examples. The improvement is more pronounced for the image-based examples, probably because graph-based examples are bottle-necked by fetching sparse data from memory rather than by the expressions themselves.

Our optimizations remove redundant bounds checks and unnecessary reads that can occur when compiling expressions that test boundary conditions. They include representing bounds as booleans, factoring the bounds out of polynomial terms, and scheduling expressions to run conditionally. They provide a significant improvement for both Shape From Shading and Image Warping. Mesh Deformation does not improve because it does not use Select.

Shape From Shading shows a significant benefit from texture use, and our register minimization heuristic provides a small benefit to Shape From Shading's JTJ function.

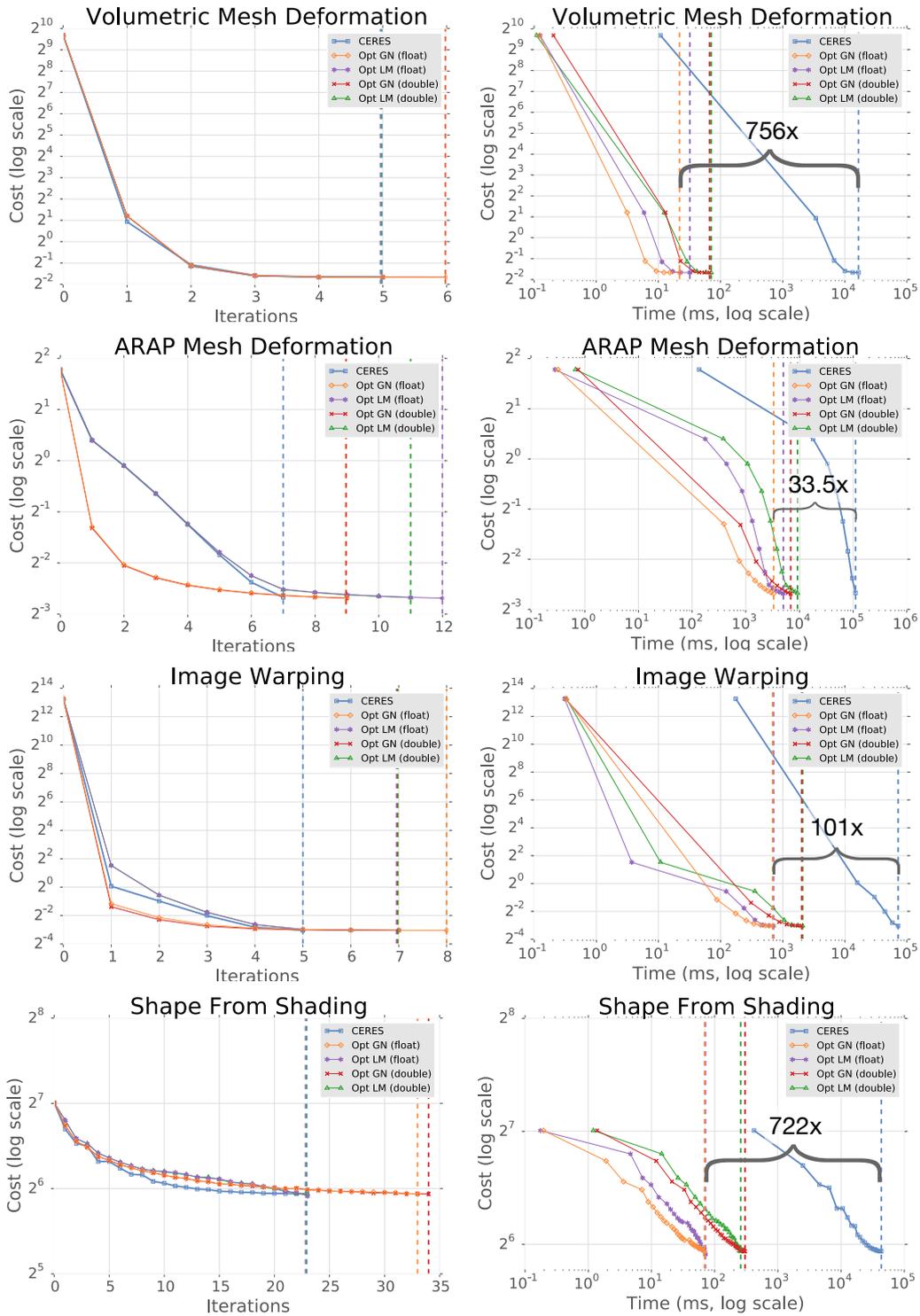

Fig. 14. Convergence of both Opt and Ceres over time, including both double/single precision and GN/LM solvers for Opt. Per non-linear iteration (left), Opt LM and Ceres converge at the similar rates, but Opt converges faster over time (right) by completing each iteration up to several orders of magnitude quicker. Cost and time are both presented using log scale, while iteration count is linear. Vertical lines are drawn for each solver type at the iteration and time when their cost dips below the final Ceres iteration. The performance gap for the fastest Opt solver variant (single-precision GN) versus Ceres is highlighted on each of the graphs on the right. Even the full double-precision Opt LM implementation (often the slowest variant) outperforms Ceres by over an order of magnitude on all problems.

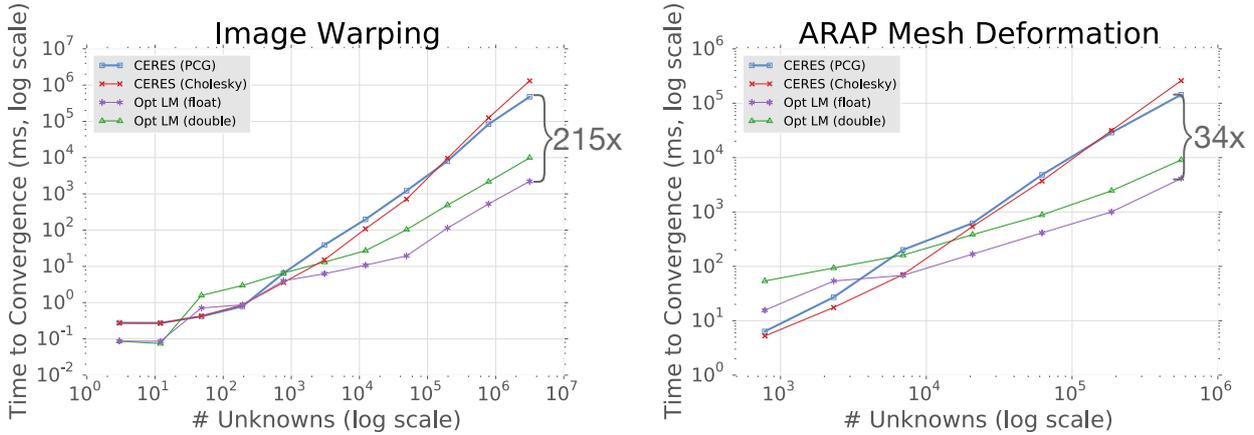

Fig. 15. Performance of Opt compared to Ceres (using a direct Cholesky solver or an iterative PCG for the linear solve) as problem size increases on two example problems. Both unknown count and time to convergence are presented in log scale. Image Warping (left), which uses implicit connectivity in Opt, has more drastic performance differences than ARAP Mesh Deformation (right). For small problems (<5k unknowns in these examples), GPU solvers are inefficient, but the Opt generated solvers rapidly become significantly faster than their Ceres equivalent as problem size increases.

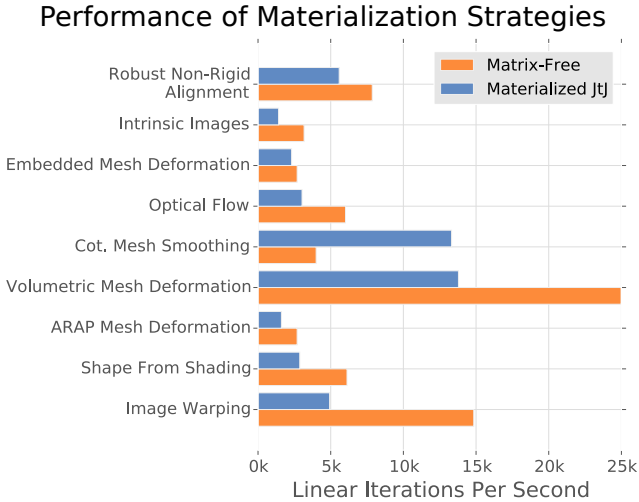

Fig. 16. Comparison of the performance between the fully materialized and the best matrix-free variant available in Opt, measuring the speed of a linear iteration in PCG. The matrix-free approach is more efficient in all cases besides Cotangent-weighted Laplacian Smoothing. The materialized version creates the $J^T J$ matrix outside of the PCG loop. This incurs a once-per-PCG-solve cost not captured by these graphs, so materialized versions will perform even worse than reported here when there are a small number of inner PCG iterations.

### 8.6 Limitations and Future Work

Currently the Opt language limits what energies can be expressed efficiently. On images, our implementation limits energies to a constant-sized neighboring stencil. However, we can extend Opt to support other neighborhood functions such as affine transformations of indices as long as the neighborhood function is invertible. We also plan to extend our graph language to support the ability to reference a variable number of neighbors (such as the edges around a vertex) to make certain energies easier to express. While some of our specific optimizations are tailored to GPUs, the overall approach of symbolically calculating and simplifying functions needed by the solver is applicable to other platforms such as multi-core CPUs, or even networked clusters of machines for large problems.

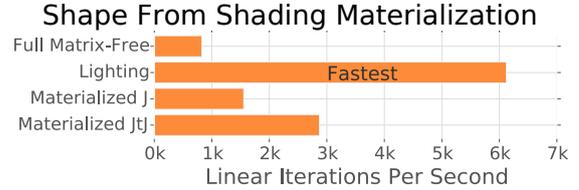

Fig. 17. Opt lets programmers specify hybrid materialization approaches, which are sometimes more efficient than either full matrix-free, or fully materialized approaches. Here, we show different strategies for the Shape From Shading example. Materializing just the lighting term (Lighting) outperforms fully matrix free, and fully materialized by 2-7x.

Finally, there are a lot of optimization problems in graphics that are not suited to the Gauss-Newton or Levenberg-Marquardt approach. Many optimization problems in the graphics literature are more efficiently solved using other techniques such as shape deformation with an interior-point optimizer [37] or mesh parametrization using quadratic programming [31]. Although these problems are not the focus of this paper, their solvers would also benefit from the architecture proposed in Opt, where a general solver library is augmented with automatically derived application-specific routines.

## 9 CONCLUSION

We have introduced Opt, a domain specific language that generates high-performance, application-specific GPU solvers from a high-level energy description based on stencils and graphs. Solvers generated with Opt are not only *orders of magnitude* faster than Ceres, but also outperform state-of-the-art hand-coded application-specific solvers which have been tuned with many month of tedious implementation effort. Opt is also highly flexible: it can generate solvers with either floating or double point precision, solvers that are matrix-free, materialized, and even intermediate hybrids, and variations of GN and LM such as IRLS or robust solvers. Further, Opt provides its own parallel PCG routines to solve for the linear intermediate systems; however, it can also hand off the linear solve to other GPU solvers such as cuSPARSE. Overall, we believe that Opt's approach of using abstracted solvers with automatically-generated application-specific routines can be extended to work with more expressive energy functions, more platforms beyond GPUs, and more kinds of solvers. Eventually, we hope that computer graphics and vision practitioners can put most energy functions from the literature into a system like Opt and automatically get a high-performance solver. We believe that Opt is a significant first step in this direction.

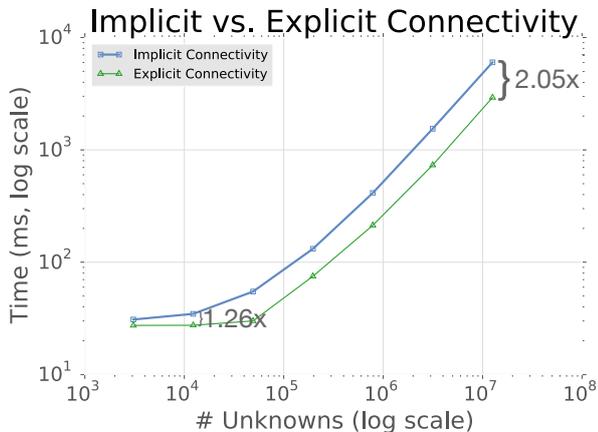

Fig. 18. Performance comparison between using the implicit connectivity of a regular grid for the Image Warping problem versus using an explicit graph representation. Both unknown count and time to convergence are presented in log scale. For medium to large size images, the explicit approach takes twice as long. Here, we configured Opt to produce Gauss-Newton solvers that run for 8 nonlinear iterations of 100 linear iterations each.

## ACKNOWLEDGMENTS

This work was supported by the DOE Office of Science ASCR in the ExMatEx and ExaCT Exascale Co-Design Centers, program manager Karen Pao; DARPA Contract No. HR0011-11-C-0007; fellowships and grants from NVIDIA, Intel, and Google; the Max Planck Center for Visual Computing and Communications, and the ERC Starting Grant 335545 CapReal; and the Stanford Pervasive Parallelism Lab (supported by Oracle, AMD, Intel, and NVIDIA). We also gratefully acknowledge hardware donations from NVIDIA Corporation.

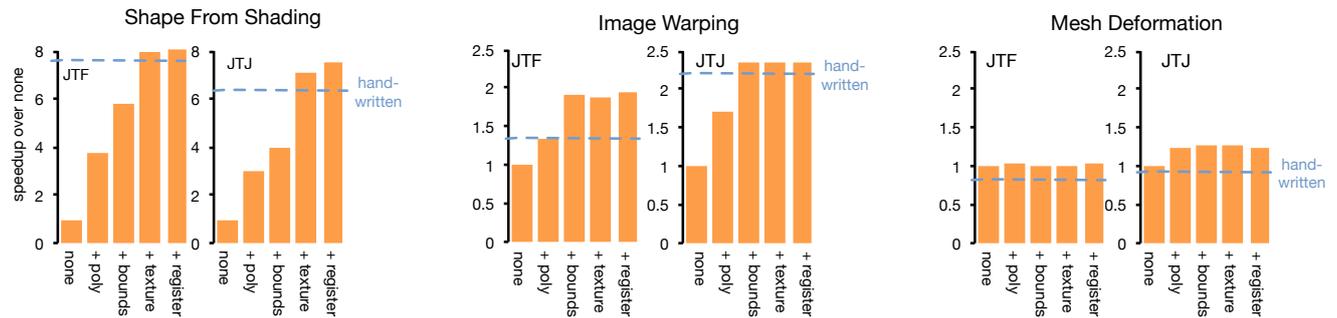

Fig. 19. The effect of our optimizations on generated code for solver routines including polynomial simplification (poly), bounds optimization (bounds), texture loads (texture), and register minimization (register), compared to a baseline with no optimization (none). The optimizations are necessary to match the performance of handwritten equivalents, and in many cases the functions run significantly faster.

## APPENDIX

In this Appendix, we provide additional details and evaluations for Opt. In Sec. A, we detail the descriptions of our test example problems that we have implemented and evaluated in Opt. The problems are split into two categories; first, those where we compare the performance against other solvers, and second, those that demonstrate the flexibility of Opt. In Sec. B, we evaluate Opt's numerical behavior on a standard optimization benchmark. We provide numbers for different solver variations generated with Opt: float vs. double and Gauss-Newton vs Levenberg-Marquardt. In Sec. C, we show how our example problems are written in Opt. We provide the Opt energies, which are similar to graphics shaders.

## A DETAILED DESCRIPTION OF EXAMPLE PROBLEMS

Along with the main contribution of Opt, we provide 10 different example optimization problems in Opt. We chose the first four example applications (see Sec. A.1) since they are commonly used in graphics research and optimized GPU code previously existed or could be easily adapted for the problem. For these problems, we implemented their energies in Opt and compare against previously written custom CUDA implementations. The hand-written CUDA baselines are (improved) versions of the authors' state-of-the-art implementations. We also implemented these applications in the Ceres solver for direct comparison to another high-level solver [1]. The remaining applications detailed in Sec. A.2 show how a variety of different applications can be solved using the Opt programming model.

### A.1 Performance Examples

*A.1.1 Image Warping.* As-rigid-as-possible Image Warping is used to interactively edit 2D shapes in a way that minimizes a warping energy. It penalizes deviations from local rigidity, while warping to a set of user-specified constraints [17]. It co-optimizes the new pixel coordinates along with the per-pixel rotation.

The CUDA implementation was adapted from the hand-written solver created by Zollhöfer et al. [73] for real-time non-rigid reconstruction. It requires around 480 lines of code to implement. Of that, 200 were devoted to the Gauss-Newton solver, and 280 to expressions for the solver routines. In this comparison, we jointly solve for rotations and translations, following the hand-written reference implementation. Note that alternating between rotation and translation in a global-local flip-flop solve is also feasible in Opt; however, overall convergence is typically worse than the joint solve [73]. In comparison, the solver generated by Opt runs about 75% faster (likely due to better bounds handling), and only requires about 20 lines of code to describe. Ceres code is more comparable in size to Opt, at around 100 lines, but it runs 100 times slower.

*A.1.2 Mesh Deformation.* As-rigid-as possible mesh deformation [50] is a variant of the previous example that shows Opt's ability to run on mesh-based problems using its graph abstraction. It defines a warping energy on the edges of the mesh rather than neighboring pixels and uses 3D coordinate frames.

The CUDA solver was also adapted from Zollhöfer et al. [73]. It is similar in size to the previous example, with around 200 lines devoted to expressing the energy, `applyJTJ`, and `evalJTF` calculations.

Opt performs around 25 times faster than a Ceres example which is implemented in around 100 lines of code.

*A.1.3 Shape From Shading.* In the Shape From Shading example, we use an optimizer to refine depth data captured by RGB-D scanners [68]. It uses a detailed color image and an estimate of the lighting based on spherical harmonics to refine the lower resolution depth information.

Shape from Shading, which is adapted from Wu et al.'s work [68], is our most complex problem. The original implementation was a patch solver variation of a Gauss Newton solver that used shared memory at the expense of per-iteration convergence. For a more direct comparison, we ported the original code into a non-patch solver, which actually improved the convergence time over the author's implementation. The CUDA code includes 445 lines to express the energy, `applyJTJ`, and `evalJTF` calculations. It took several months for a group of researchers to implement and optimize. In comparison, the Opt solver code is around 100 lines and runs more than 30% faster. Some of this improvement is due to using texture objects to represent the images, which is an optimization that the original authors did not have time to do.

Shape from Shading also benefits from using pre-computed arrays. We instruct Opt to pre-compute a lighting term and a boundary term that are expensive to calculate and used by the energy of multiple pixels. Without this annotation, Opt runs over 7 times slower. We expect that other complicated problems will have similar behavior and pre-computed arrays will give the user an easy way to experiment with how the computation is scheduled.

*A.1.4 Poisson Image Editing.* Poisson Image Editing is used to splice a source image into a target image without introducing seams [47]. Its energy function preserves the gradients of the source image while matching the boundary to gradients in the target image. The energy formulation in this function makes $J$ constant, and our matrix-free solver is able to inline those constants into the solver routines rather than load them from memory.

To compare against a CUDA version, we adapted the Image Warping CUDA example to use the Poisson Image Editing objective function, which uses about 67 lines for the energy, `evalJTF`, and `applyJTJ`. Opt performs about 10% faster and uses only about 15 lines of code. Since this problem boils down to solving a linear system of equations, we also compare against Eigen [25], a high-performance linear-algebra library for CPUs using Cholesky with pre-ordering since it was fastest. The entire Opt solve was 50 times faster than Eigen's matrix solve (not including its matrix setup time), due to Opt's ability to implicitly represent the connectivity of the matrix.

### A.2 Expressiveness Examples

Our performance evaluations (see Sec. A.1) focus on examples from the literature where state-of-the-art hand-written code previously existed and can be compared. Opt's programming model is also able to handle a wider variety of *general* non-linear least squares problems, which is at the core of many computer graphics and vision problems.

*A.2.1 Embedded Deformation.* Embedded Deformation is a popular alternative method to as-rigid-as-possible deformation [51]. Rather than solving for a per-vertex rotation parameterized by Euler angles, Embedded Deformation solves for a full transformation matrix and enforces the rotations via additional soft-constraints. Compared to writing a solver by hand, writing Embedded Deformation in Opt was an easy process, since it only required increasing the number of unknowns and changing the energy terms of our as-rigid-as-possible energy, which amounted to tens of lines of code and under an hour of work. Gauss-Newton solvers are fragile on this energy function, so we configured Opt to generate an LM solver instead. The solver it produces can deform a 12k vertex mesh at an interactive rate of 5 frames per second and only requires an energy function of around 40 lines.

*A.2.2 Cotangent-weighted Laplacian Smoothing.* Cotangent-weighted Laplacian Smoothing is a method for smoothing meshes that tries to preserves the area of triangles adjacent to each edge [15]. It adapts well to meshes with non-uniform tessellations. This example highlights Opt's ability to define residuals on larger components of a mesh by defining a hyper-edge in our graph representation that contains all the vertices in a wedge at each edge. We show the power of Opt by implementing a small variant that allows the cotangent weights to be recomputed during deformation instead of using the values from the original mesh. This variant is normally hard to write since it introduces complicated derivative terms. Opt generates them automatically, making it easier to experiment with small variants on existing energy functions. Opt generates a solver from around 45 lines of code that can smooth a 44k vertex mesh at 3.3 fps.

*A.2.3 Dense Optical Flow.* Dense Optical Flow computes the apparent motion of objects between frames in a video at the pixel level. We implement a hierarchical version of Horn and Schunck's algorithm [29] using an iterative relaxation scheme. Because optical flow is searching for correspondences in images, its unknowns are used to sample values from the input frames. We support this pattern using a *sampled image* operator, which can be accessed with arbitrary $(u, v)$ coordinates. When these coordinates are dependent on the unknown image, the user provides the directional derivatives of the sampled image as other input images, which will be used to lookup the partials for the operator in the symbolic differentiation. The solver Opt generates from around 20 lines of code solves optical flow at 5.38MP/s.

*A.2.4 Intrinsic Image Decomposition.* Intrinsic Image Decomposition separates an input image into its reflectance and shading components. This problem dates back to the seminal Retinex [35] work, which assumes that large gradients are more likely to correspond to reflectance than shading variation. Current approaches [8, 41] implement this assumption based on energy minimization by enforcing reflectance sparseness and shading smoothness. We reimplemented a basic version of Meka et al. [41] in Opt to show its ability to solve optimization problems that involve sparsity inducing norms; see Fig. 20. Specifying the three objectives (reproduction of input image, $\ell_p$ reflectance sparsity and $\ell_2$ shading smoothness) only requires 37 lines of code in the Opt language. Opt solves the intrinsic image decomposition problem for $461k$ unknowns (pixel resolution of $640 \times 360$, 2 unknowns per pixel), in less than 25.1 ms, fast enough to operate in real time on 30Hz video. Internally, Opt minimizes the energy based on Iteratively Reweighted Least Squares (IRLS).

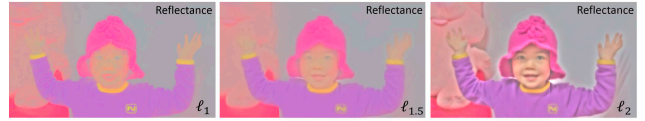

Fig. 20. Opt easily allows exploration of different $\ell_p$-norms by generating IRLS solvers. In Meka et al. [41], smaller values of $p$ better separate reflectance and shading.

*A.2.5 Volumetric Mesh Deformation.* Volumetric Mesh Deformation is an alternative approach to mesh deformation that warps an underlying 3-dimensional grid that the mesh is embedded in; e.g., see Innmann et al. [30]. The structure of the problem is implicit in the grid representation, so we stand to gain more than standard graph problems by using Opt over a solver that materializes the intermediate matrices. To demonstrate this performance benefit, we implemented this problem in both Ceres and Opt. The solver Opt generates from 27 lines of code solves for the deformation of a 4961 voxel grid at 21.4ms, sufficient for real-time applications. This is more than 760x faster than the Ceres solver.

*A.2.6 Robust Non-Rigid Alignment.* Robust optimization is an alternative to robust norms that is often used for computer vision problems. The idea is to introduce auxiliary variables that weight data points as part of the optimization process. For instance, this strategy is often used in bundle adjustment or non-rigid deformation frameworks to determine the reliability of correspondences [38, 58, 71, 73]. In Opt, it is easy to add these terms using additional unknowns for energy functions on single and mixed domains. We implemented a robust term along with a fitting term for our Mesh Deformation example in only 3 extra lines of solver code; the solver it produces can deform a 10k vertex mesh at an interactive rate of 3.7 frames per second. In this example, we include a data fitting term that constrains the deformation based on a target point cloud using a point-to-plane term; i.e., a non-rigid ICP. As ground truth 3D captures, we use the data from Vlasic et al. [59, 60]. In order to make the problem harder, we introduce artificial noise by adding spurious correspondences that are far off from the surface. During the optimization, the robust optimization minimizes the weight of these outliers, and is able to achieve a robust non-rigid alignment of the mesh with respect to the target point cloud. A visual comparison with and without robust optimization is shown in Fig. 21, bottom right.

## B NUMERIC EVALUATION

In this section, we provide a numeric evaluation of Opt. Our aim is not to evaluate performance, but rather to exercise numerics. To this end, we implemented all univariate energies from the National Institute of Standards and Technology (NIST) benchmark for non-linear least squares problems[2], a standardized benchmark for optimization. These objectives are non-linear regression problems that contain few unknowns (2-9) each with up to a few hundred data points; they are classified by three levels of difficulty (lower, average, higher). On all of these problems, we run four Opt solvers, float vs. double and GN vs. LM, as well as the Ceres solver which uses LM with double precision. The results are visualized in Fig. 22 and Fig. 23. As we can see there is virtually no difference between the solver types on easier problems, on some of the harder problems combinations of the numerically more stable double precision and LM are necessary – our reference is to match Ceres.

## C   EXAMPLE ENERGY SPECIFICATIONS

Fig. 24 through Fig. 33 list the Opt code for each of the ten energy functions described in Section A. Opt code is based on the Lua programming language, but uses specialized operations to specify the dimensions and types of unknowns and other variables. All energy functions are centered at a local origin, with all access being relative to this origin, which is replicated over the domain of each energy function.

---

[2]http://www.itl.nist.gov/div898/strd/nls/nls_main.shtml

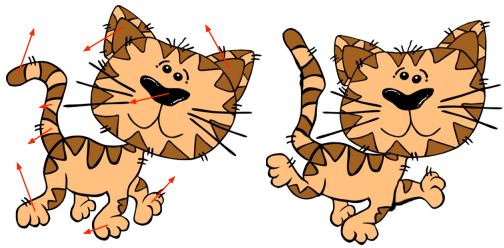

*Image Warping*   539k unknowns, 3 unknowns per pixel, 5-point stencil
0.250 MP/s, 1.8x faster than CUDA, 101x Ceres

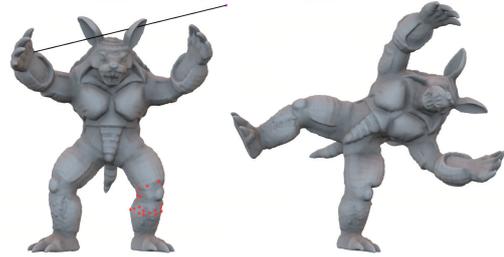

*ARAP Mesh Deformation*   360k unknowns, 6 per vertex, vertex and edge energies
18.1 kverts/s, 1.6x faster than CUDA, 33.6x Ceres

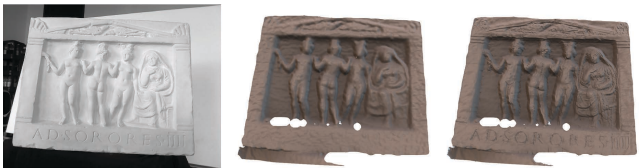

*Shape From Shading*   192k unknowns, 1 per pixel, 9-point stencil
2.99 MP/s, 1.3x faster than CUDA, 722x Ceres

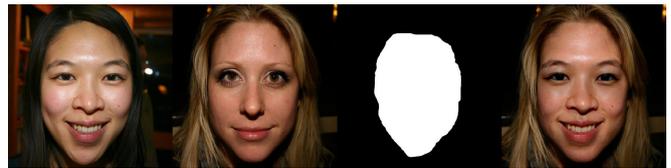

*Poisson Image Editing*   1.3M unknowns, 4 unknowns per pixel, 5-point stencil
26.3 MP/s 1.1x faster than CUDA, 50.0x Eigen

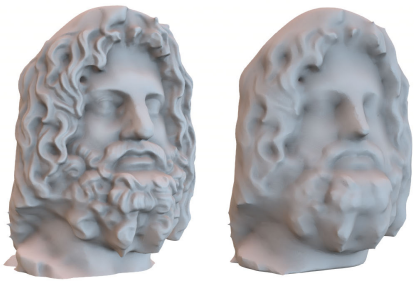

*Cotangent Mesh Smoothing*   132k unknowns, 3 per vertex,
triangle and vertex energies, 147 kverts/s

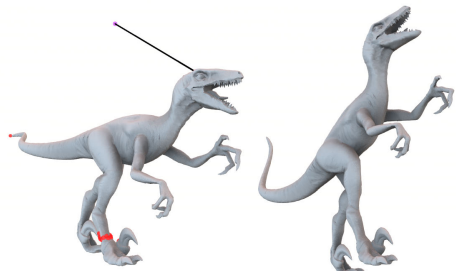

*Embedded Mesh Deformation*   154k unknowns, 12 per vertex, vertex and edge energies
66.4 kverts/s

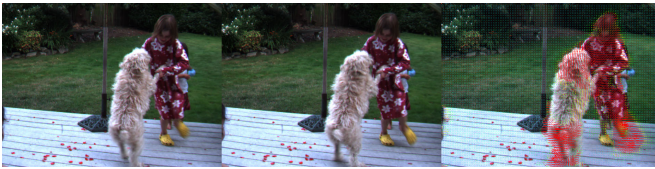

*Optical Flow*   614k unknowns, 2 per pixel, 5-point stencil
5.38 MP/s

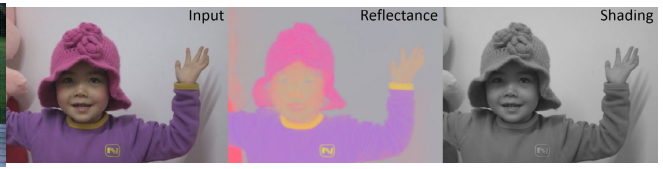

*Intrinsic Image Decomposition*   461k unknowns, 2 per pixel, 5-point stencil
9.19 MP/s

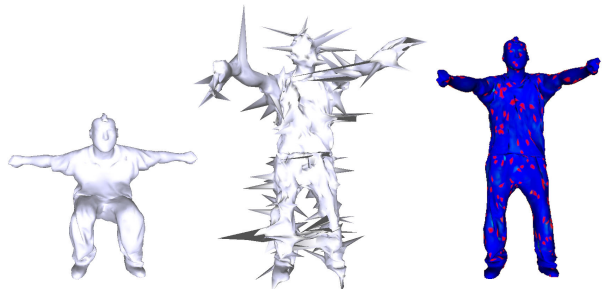

Starting Mesh   Non-robust with outliers   Robust with outliers
*Robust Non-Rigid Alignment*   70k unknowns, 7 per vertex, vertex and edge energies
37.1 kverts/s

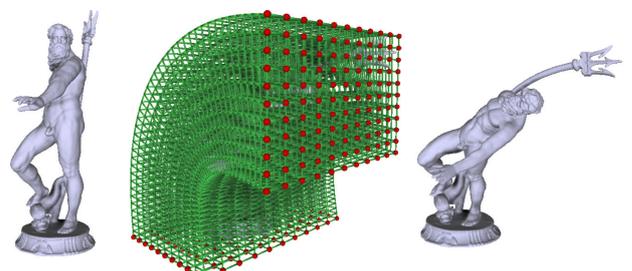

*Volumetric Mesh Deformation*   30k unknowns, 7-point stencil (3D)
232 kvoxels/s, 769x faster than Ceres

Fig. 21. Visualization of our example problems implemented in Opt.

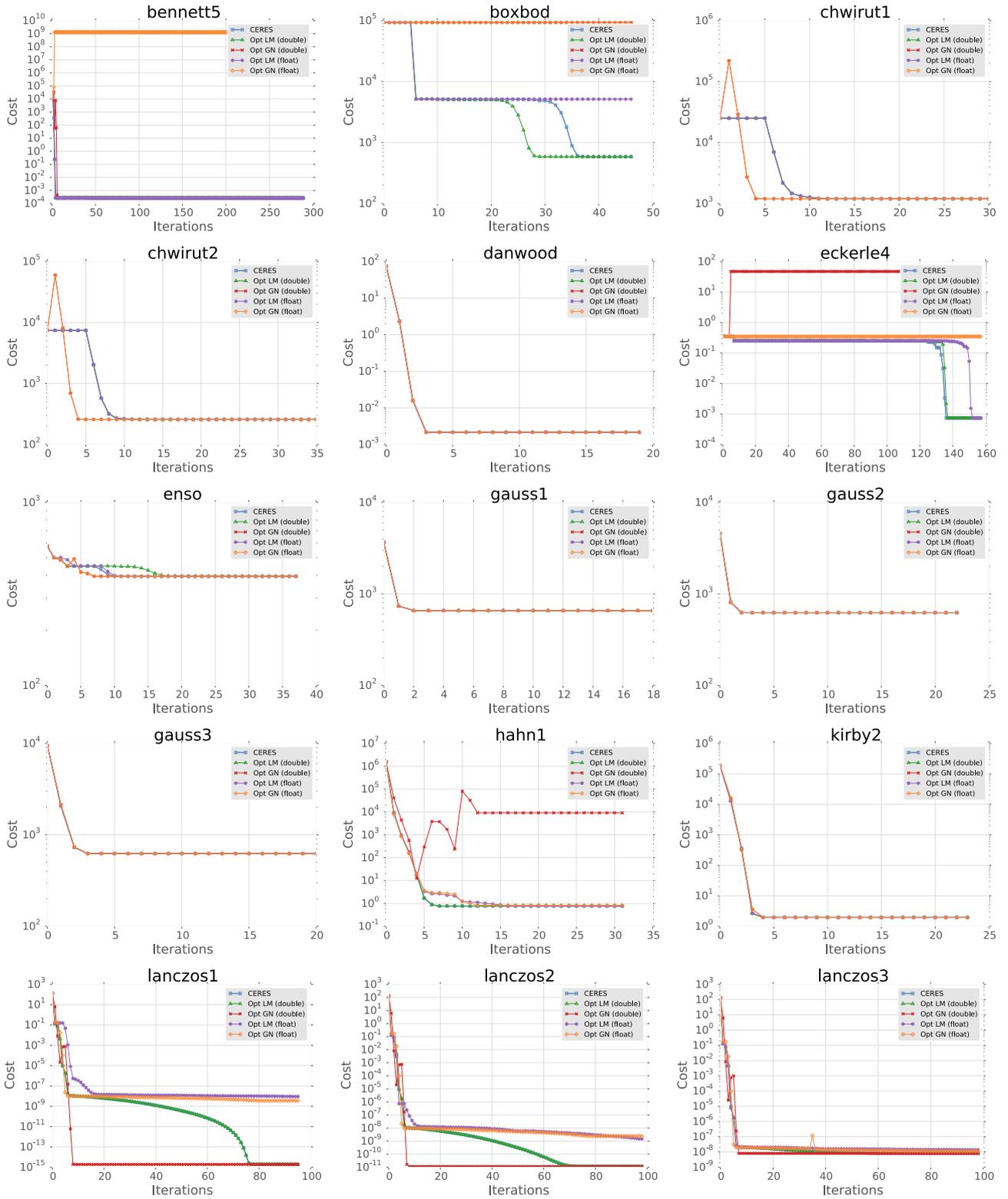

Fig. 22. Cost vs. iterations on the first 15 univariate problems from the NIST non-linear least squares optimization benchmark: http://www.itl.nist.gov/div898/strd/nls/nls_main.shtml.

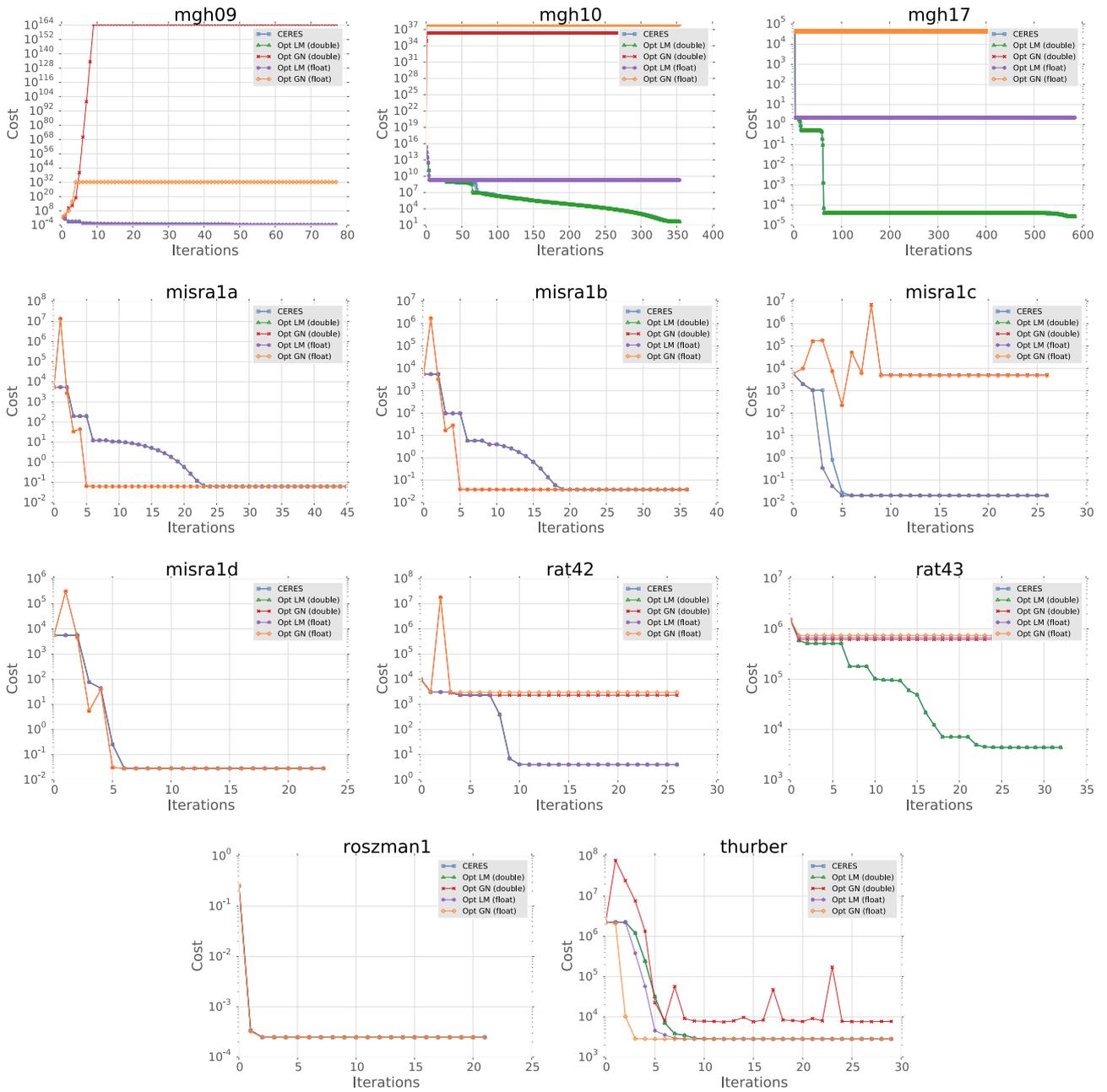

Fig. 23. Cost vs. iterations on the last 11 univariate problems from the NIST non-linear least squares optimization benchmark: http://www.itl.nist.gov/div898/strd/nls/nls_main.shtml.

```
local W,H = Dim("W",0), Dim("H",1)
local Offset = Unknown("Offset",float2,{W,H},0)
local Angle = Unknown("Angle",float,{W,H},1)
local UrShape = Array("UrShape", float2,{W,H},2) --original mesh position
local Constraints = Array("Constraints", float2,{W,H},3) -- user constraints
local Mask = Array("Mask", float, {W,H},4) -- validity mask for mesh
local w_fitSqrt = Param("w_fitSqrt", float, 5)
local w_regSqrt = Param("w_regSqrt", float, 6)
Exclude(Not(eq(Mask(0,0),0)))

--regularization
for x,y in Stencil { {1,0}, {-1,0}, {0,1}, {0, -1} } do
    local e_reg = w_regSqrt*((Offset(0,0) - Offset(x,y))
                    - Rotate2D(Angle(0,0),(UrShape(0,0) - UrShape(x,y))))
    local valid = InBounds(x,y) * eq(Mask(x,y),0) * eq(Mask(0,0),0)
    Energy(Select(valid,e_reg,0))
end
--fitting
local e_fit = (Offset(0,0)- Constraints(0,0))
local valid = All(greatereq(Constraints(0,0),0))
Energy(w_fitSqrt*Select(valid, e_fit , 0.0))
```

Fig. 24. Image Warping energy function in Opt.

```
local N = opt.Dim("N",0)
local w_fitSqrt =  Param("w_fitSqrt", float, 0)
local w_regSqrt =  Param("w_regSqrt", float, 1)
local Offset =     Unknown("Offset", float3,{N},2)          --vertex.xyz, rotation.xyz <- unknown
local Angle =      Unknown("Angle",float3,{N},3)
local UrShape =    Array("UrShape",float3,{N},4)            --original position: vertex.xyz
local Constraints = Array("Constraints",float3,{N},5)       --user constraints
local G = Graph("G", 6, "v0", {N}, 7, "v1", {N}, 9)
UsePreconditioner(true)

--fitting
local e_fit = Offset(0) - Constraints(0)
local valid = greatereq(Constraints(0,0), -999999.9)
Energy(Select(valid,w_fitSqrt*e_fit,0))

--regularization
local ARAPCost = (Offset(G.v0) - Offset(G.v1)) - Rotate3D(Angle(G.v0),UrShape(G.v0) - UrShape(G.v1))
Energy(w_regSqrt*ARAPCost)
```

Fig. 25. ARAP Mesh Deformation energy function in Opt.

```
local DEPTH_DISCONTINUITY_THRE = 0.01
local W,H       = Dim("W",0), Dim("H",1)
local w_p       = sqrt(Param("w_p",float,0))-- Fitting weight
local w_s       = sqrt(Param("w_s",float,1))-- Regularization weight
local w_g       = sqrt(Param("w_g",float,2))-- Shading weight
local f_x       = Param("f_x",float,3)
local f_y       = Param("f_y",float,4)
local u_x       = Param("u_x",float,5)
local u_y       = Param("u_y",float,6)
local L = {}
for i=1,9 do L[i] = Param("L_" .. i .. "",float,6+i) end -- lighting model parameters
local X         = Unknown("X",opt_float, {W,H},16) -- Refined Depth
local D_i       = Array("D_i",opt_float, {W,H},17) -- Depth input
local Im        = Array("Im",opt_float, {W,H},18) -- Target Intensity
local edgeMaskR = Array("edgeMaskR",uint8, {W,H},19) -- Edge mask.
local edgeMaskC = Array("edgeMaskC",uint8, {W,H},20) -- Edge mask.
local posX,posY = Index(0),Index(1)
-- equation 8
function p(offX,offY)
    local d = X(offX,offY)
    local i = offX + posX
    local j = offY + posY
    return Vector(((i-u_x)/f_x)*d, ((j-u_y)/f_y)*d, d)
end
-- equation 10
function normalAt(offX, offY)
        local i = offX + posX
    local j = offY + posY
    local n_x = X(offX, offY - 1) * (X(offX, offY) - X(offX - 1, offY)) / f_y
    local n_y = X(offX - 1, offY) * (X(offX, offY) - X(offX, offY - 1)) / f_x
    local n_z = (n_x * (u_x - i) / f_x) + (n_y * (u_y - j) / f_y) - (X(offX-1, offY)*X(offX, offY-1) / (f_x*f_y))
    local sqLength = n_x*n_x + n_y*n_y + n_z*n_z
    local inverseMagnitude = Select(greater(sqLength, 0.0), 1.0/sqrt(sqLength), 1.0)
    return inverseMagnitude * Vector(n_x, n_y, n_z)
end
function B(offX, offY)
        local normal = normalAt(offX, offY)
        local n_x = normal[0]
        local n_y = normal[1]
        local n_z = normal[2]
        return L[1] +
        L[2]*n_y + L[3]*n_z + L[4]*n_x   +
        L[5]*n_x*n_y + L[6]*n_y*n_z + L[7]*(-n_x*n_x - n_y*n_y + 2*n_z*n_z) + L[8]*n_z*n_x + L[9]*(n_x*n_x-n_y*n_y)
end
function I(offX, offY)
        return Im(offX,offY)*0.5 + 0.25*(Im(offX-1,offY)+Im(offX,offY-1))
end
local function DepthValid(x,y) return greater(D_i(x,y),0) end
local function B_I(x,y)
    local bi = B(x,y) - I(x,y)
    local valid = DepthValid(x-1,y)*DepthValid(x,y)*DepthValid(x,y-1)
    return Select(InBoundsExpanded(0,0,1)*valid,bi,0)
end
B_I = ComputedArray("B_I", {W,H}, B_I(0,0))
-- do not include unknowns for where the depth is invalid
Exclude(Not(DepthValid(0,0)))
-- fitting term
local E_p = X(0,0) - D_i(0,0)
Energy(Select(DepthValid(0,0),w_p*E_p,0))
-- shading term
local E_g_h = (B_I(0,0) - B_I(1,0))*edgeMaskR(0,0)
local E_g_v = (B_I(0,0) - B_I(0,1))*edgeMaskC(0,0)
Energy(Select(InBoundsExpanded(0,0,1),w_g*E_g_h,0))
Energy(Select(InBoundsExpanded(0,0,1),w_g*E_g_v,0))
-- regularization term
local function Continuous(x,y) return less(abs(X(0,0) - X(x,y)),DEPTH_DISCONTINUITY_THRE) end
local valid = DepthValid(0,0)*DepthValid(0,-1)*DepthValid(0,1)*DepthValid(-1,0)*DepthValid(1,0)*
                    Continuous(0,-1)*Continuous(0,1)*Continuous(-1,0)*Continuous(1,0)*InBoundsExpanded(0,0,1)
local validArray = ComputedArray("valid", {W,H},valid)
valid = eq(validArray(0,0),1)
local E_s = 4.0*p(0,0) - (p(-1,0) + p(0,-1) + p(1,0) + p(0,1))
Energy(Select(valid,w_s*E_s,0))
```

Fig. 26. Shape from Shading energy function in Opt.

```lua
local W,H = Dim("W",0), Dim("H",1)
local X = Unknown("X", float4,{W,H},0) -- unknown, initialized to base image
local T = Array("T", float4,{W,H},1) -- inserted image
local M = Array("M", float, {W,H},2) -- mask, excludes parts of base image
UsePreconditioner(false)

-- do not include unmasked pixels in the solve
Exclude(Not(eq(M(0,0),0)))

for x,y in Stencil { {1,0},{-1,0},{0,1},{0,-1} } do
    local e = (X(0,0) - X(x,y)) - (T(0,0) - T(x,y))
    Energy(Select(InBounds(x,y),e,0))
end
```

Fig. 27. Poisson Image Editing energy function in Opt.

```lua
local N = Dim("N",0)

local w_fitSqrt = Param("w_fitSqrt", float, 0)
local w_regSqrt = Param("w_regSqrt", float, 1)
local w_rotSqrt = Param("w_rotSqrt", float, 2)
local X =                 Unknown("X", float12,{N},3)            --vertex.xyz, rotation_matrix <- unknown
local UrShape =           Image("UrShape", float3,{N},4)         --urshape: vertex.xyz
local Constraints = Image("Constraints", float3,{N},5) --constraints
local G = Graph("G", 6, "v0", {N}, 7, "v1", {N}, 9)
UsePreconditioner(true) --really needed here

local Offset = Slice(X,0,3) -- select part of unknown for position

--fitting
local e_fit = Offset(0) - Constraints(0)
local valid = greatereq(Constraints(0)(0), -999999.9)
Energy(Select(valid, w_fitSqrt*e_fit, 0))

--rot
local RotMatrix = Slice(X,3,12) -- extract rotation matrix
local R = RotMatrix(0)
local c0 = Vector(R(0), R(3), R(6))
local c1 = Vector(R(1), R(4), R(7))
local c2 = Vector(R(2), R(5), R(8))
Energy(w_rotSqrt*Dot(c0,c1))
Energy(w_rotSqrt*Dot(c0,c2))
Energy(w_rotSqrt*Dot(c1,c2))
Energy(w_rotSqrt*(Dot(c0,c0)-1))
Energy(w_rotSqrt*(Dot(c1,c1)-1))
Energy(w_rotSqrt*(Dot(c2,c2)-1))

local regCost = (Offset(G.v1) - Offset(G.v0)) -
                Matrix3x3Mul(RotMatrix(G.v0), (UrShape(G.v1) - UrShape(G.v0)))
Energy(w_regSqrt*regCost)
```

Fig. 28. Embedded Mesh Deformation energy function in Opt.

```lua
local N = Dim("N",0)

local w_fitSqrt = Param("w_fit", float, 0)
local w_regParam = Param("w_reg", float, 1)
local X = Unknown("X", float3,{N},2)
local A = Array("A", float3,{N},3)
local G = Graph("G", 4, "v0", {N}, 5, --current vertex
                              "v1", {N}, 7, --neighboring vertex
                              "v2", {N}, 9, --prev neighboring vertex
                              "v3", {N}, 11) --next neighboring vertex

UsePreconditioner(true)

function cot(v0, v1)
        local adotb = Dot3(v0, v1)
        local disc = Dot3(v0, v0)*Dot3(v1, v1) - adotb*adotb
        disc = Select(greater(disc, 0.0), disc,  0.0001)
        return Dot3(v0, v1) / Sqrt(disc)
end

-- fit energy
Energy(w_fitSqrt*(X(0) - A(0)))

local a = normalize(X(G.v0) - X(G.v2)) --float3
local b = normalize(X(G.v1) - X(G.v2)) --float3
local c = normalize(X(G.v0) - X(G.v3)) --float3
local d = normalize(X(G.v1) - X(G.v3)) --float3

--cotangent laplacian; Meyer et al. 03
local w = 0.5*(cot(a,b) + cot(c,d))
w = Sqrt(Select(greater(w, 0.0), w, 0.0001))
Energy(w_regSqrt*w*(X(G.v1) - X(G.v0)))
```

Fig. 29.  Cotangent Mesh Smoothing energy function in Opt.

```lua
local W,H = Dim("W",0), Dim("H",1)
local w_fitSqrt = Param("w_fit", float, 0)
local w_regSqrt = Param("w_reg", float, 1)
local X = Unknown("X", float2,{W,H},2)
local I = Array("I",float,{W,H},3)
local I_hat_im = Array("I_hat",float,{W,H},4)
local I_hat_dx = Array("I_hat_dx",float,{W,H},5)
local I_hat_dy = Array("I_hat_dy",float,{W,H},6)
local I_hat = SampledImage(I_hat_im,I_hat_dx,I_hat_dy)

local i,j = Index(0), Index(1)
UsePreconditioner(false)
-- fitting
local e_fit = w_fitSqrt*(I(0,0) - I_hat(i + X(0,0,0),j + X(0,0,1)))
Energy(e_fit)
-- regularization
for nx,ny in Stencil { {1,0}, {-1,0}, {0,1}, {0,-1} } do
        local e_reg = w_regSqrt*(X(0,0) - X(nx,ny))
        Energy(Select(InBounds(nx,ny),e_reg,0))
end
```

Fig. 30.  Optical Flow energy function in Opt.

```
local W,H,D = Dim("W",0), Dim("H",1), Dim("D",2)

local Offset =        Unknown("Offset", float3,{W,H,D},0)      --vertex.xyz, rotation.xyz <- unknown
local Angle =         Unknown("Angle",float3,{W,H,D},1)
local UrShape =       Array("UrShape",float3,{W,H,D},2)        --original position: vertex.xyz
local Constraints = Array("Constraints",float3,{W,H,D},3)      --user constraints
local w_fitSqrt =   Param("w_fitSqrt", float, 4)
local w_regSqrt =   Param("w_regSqrt", float, 5)
UsePreconditioner(true)

--fitting
local e_fit = Offset(0,0,0) - Constraints(0,0,0)
local valid = greatereq(Constraints(0,0,0)(0), -999999.9)
Energy(Select(valid,w_fitSqrt*e_fit,0))

for i,j,k in Stencil { {1,0,0}, {-1,0,0}, {0,1,0}, {0,-1,0}, {0,0,1}, {0,0,-1}} do
        local ARAPCost = (Offset(0,0,0) - Offset(i,j,k)) - Rotate3D(Angle(0,0,0),UrShape(0,0,0) - UrShape(i,j,k))
        local ARAPCostF = Select(InBounds(0,0,0),        Select(InBounds(i,j,k), ARAPCost, 0.0), 0.0)
        Energy(w_regSqrt*ARAPCostF)
end
```

Fig. 31. Volumetric Mesh Deformation energy function in Opt.

```
local N = Dim("N",0)
local w_fitSqrt        = Param("w_fitSqrt", float, 0)
local w_regSqrt        = Param("w_regSqrt", float, 1)
local w_confSqrt       = 0.1
local Offset           = Unknown("Offset", float3,{N},2)           --vertex.xyz, rotation.xyz <- unknown
local Angle            = Unknown("Angle", float3,{N},3)            --vertex.xyz, rotation.xyz <- unknown
local RobustWeights    = Unknown("RobustWeights", float,{N},4)
local UrShape          = Array("UrShape", float3, {N},5)           --urshape: vertex.xyz
local Constraints      = Array("Constraints", float3,{N},6)        --constraints
local ConstraintNormals = Array("ConstraintNormals", float3,{N},7)
local G = Graph("G", 8, "v0", {N}, 9, "v1", {N}, 10)
UsePreconditioner(true)

local robustWeight = RobustWeights(0)
--fitting
local e_fit = robustWeight*ConstraintNormals(0):dot(Offset(0) - Constraints(0))
local validConstraint = greatereq(Constraints(0), -999999.9)
Energy(w_fitSqrt*Select(validConstraint, e_fit, 0.0))

--RobustWeight Penalty
local e_conf = 1-(robustWeight*robustWeight)
e_conf = Select(validConstraint, e_conf, 0.0)
Energy(w_confSqrt*e_conf)

--regularization
local ARAPCost = (Offset(G.v0) - Offset(G.v1)) - Rotate3D(Angle(G.v0),UrShape(G.v0) - UrShape(G.v1))
Energy(w_regSqrt*ARAPCost)
```

Fig. 32. Robust Mesh Deformation energy function in Opt.

```
W,H = opt.Dim("W",0), opt.Dim("H",1)
local w_fitSqrt           = Param("w_fitSqrt", float, 0)
local w_regSqrtAlbedo     = Param("w_regSqrtAlbedo", float, 1)
local w_regSqrtShading    = Param("w_regSqrtShading", float, 2)
local pNorm               = Param("pNorm", float, 4)
local r                   = Unknown("r", float3,{W,H},5)
local r_const             = Array("r_const", float3,{W,H},5) -- A constant view of the unknown
local i                   = Array("i", float3,{W,H},6)
local s                   = Unknown("s", float,{W,H},8)

-- reg Albedo
for x,y in Stencil { {1,0}, {-1,0}, {0,1}, {0,-1} } do
        local diff = (r(0,0) - r(x,y))
        local diff_const = (r_const(0,0) - r_const(x,y))
        -- The helper L_p function takes diff_const, raises it's length to the (p-2) power,
        -- and stores it in a computed array, so its value remains constant during the nonlinear iteration,
        -- then multiplies it with diff and returns
        local laplacianCost = L_p(diff, diff_const, pNorm, {W,H})
        local laplacianCostF = Select(InBounds(0,0),Select(InBounds(x,y), laplacianCost,0),0)
        Energy(w_regSqrtAlbedo*laplacianCostF)
end

-- reg Shading
for x,y in Stencil { {1,0}, {-1,0}, {0,1}, {0,-1} } do
        local diff = (s(0,0) - s(x,y))
        local laplacianCostF = Select(InBounds(0,0),Select(InBounds(x,y), diff,0),0)
        Energy(w_regSqrtShading*laplacianCostF)
end

-- fit
local fittingCost = r(0,0)+s(0,0)-i(0,0)
Energy(w_fitSqrt*fittingCost)
```

Fig. 33.  Intrinsic Image Decomposition energy function in Opt.